\documentclass[longauth]{aa}

\usepackage[utf8]{inputenc}
\usepackage{txfonts}
\usepackage{graphicx}
\usepackage{sidecap}
\usepackage[colorinlistoftodos]{todonotes}
\usepackage[colorlinks=true, allcolors=blue]{hyperref}
\usepackage{enumitem}
\usepackage{times}
\usepackage{bm}
\usepackage{graphicx}
\usepackage{euclid}
\usepackage{soul}

\newcommand{\de}{\mathrm{DE}}

\defcitealias{Euclid_IST}{EC20}

\title{\Euclid preparation: XV. Forecasting cosmological constraints for the \Euclid and CMB joint analysis}

\titlerunning{\Euclid preparation: XV. Forecasting cosmological constraints for the \Euclid and CMB joint analysis}

\author{Euclid Collaboration: S.~Ili\'c$^{1,2,3}$\thanks{\email{stephane.ilic@obspm.fr}}, N.~Aghanim$^{4}$, C.~Baccigalupi$^{5,6,7,8}$, J.R.~Bermejo-Climent$^{9,10,11}$, G.~Fabbian$^{12,114}$, L.~Legrand$^{4,13}$, D.~Paoletti$^{10,14}$, M.~Ballardini$^{10,11,15}$, M.~Archidiacono$^{16,17}$, M.~Douspis$^{4}$
% ...
}

\authorrunning{Euclid Collaboration}

\institute{$^{1}$ Universit\'e PSL, Observatoire de Paris, Sorbonne Universit\'e, CNRS, LERMA, F-75014, Paris, France\\
$^{2}$ CEICO, Institute of Physics of the Czech Academy of Sciences, Na Slovance 2, Praha 8, Czech Republic\\
$^{3}$ Institut de Recherche en Astrophysique et Plan\'etologie (IRAP), Universit\'e de Toulouse, CNRS, UPS, CNES, 14 Av. Edouard Belin, F-31400 Toulouse, France\\
$^{4}$ Universit\'e Paris-Saclay, CNRS, Institut d'astrophysique spatiale, 91405, Orsay, France\\
$^{5}$ IFPU, Institute for Fundamental Physics of the Universe, via Beirut 2, 34151 Trieste, Italy\\
$^{6}$ SISSA, International School for Advanced Studies, Via Bonomea 265, I-34136 Trieste TS, Italy\\
$^{7}$ INFN, Sezione di Trieste, Via Valerio 2, I-34127 Trieste TS, Italy\\
$^{8}$ INAF-Osservatorio Astronomico di Trieste, Via G. B. Tiepolo 11, I-34131 Trieste, Italy\\
$^{9}$ Departamento de Astrof\'{i}sica, Universidad de La Laguna, E-38206, La Laguna, Tenerife, Spain\\
$^{10}$ INAF-Osservatorio di Astrofisica e Scienza dello Spazio di Bologna, Via Piero Gobetti 93/3, I-40129 Bologna, Italy\\
$^{11}$ INFN-Bologna, Via Irnerio 46, I-40126 Bologna, Italy\\
$^{12}$ Center for Computational Astrophysics, Flatiron Institute, 162 5th Avenue, 10010, New York, NY, USA \\
$^{13}$ Universit\'e de Gen\`eve, D\'epartement de Physique Th\'eorique and Centre for Astroparticle Physics, 24 quai Ernest-Ansermet, CH-1211 Gen\`eve 4, Switzerland\\
$^{14}$ INFN-Sezione di Bologna, Viale Berti Pichat 6/2, I-40127 Bologna, Italy\\
$^{15}$ Dipartimento di Fisica e Astronomia, Universit\'a di Bologna, Via Gobetti 93/2, I-40129 Bologna, Italy\\
$^{16}$ Dipartimento di Fisica `Aldo Pontremoli', Universit\'a degli Studi di Milano, Via Celoria 16, I-20133 Milano, Italy\\
$^{17}$ INFN-Sezione di Milano, Via Celoria 16, I-20133 Milano, Italy\\
$^{18}$ Universit\'e de Paris, CNRS, Astroparticule et Cosmologie, F-75013 Paris, France\\
$^{19}$ Instituto de Astrof\'{i}sica de Canarias, Calle V\'{i}a L\`actea s/n, 38204, San Crist\`obal de la Laguna, Tenerife, Spain\\
$^{20}$ Centro de Estudios de F\'isica del Cosmos de Arag\'on (CEFCA), Plaza San Juan, 1, planta 2, E-44001, Teruel, Spain\\
$^{21}$ Istituto Nazionale di Fisica Nucleare, Sezione di Ferrara, Via Giuseppe Saragat 1, I-44122 Ferrara, Italy\\
$^{22}$ Dipartimento di Fisica, Universit\`a di Roma Tor Vergata, Via della Ricerca Scientifica 1, Roma, Italy\\
$^{23}$ INFN, Sezione di Roma 2, Via della Ricerca Scientifica 1, Roma, Italy\\
$^{24}$ INAF-IASF Milano, Via Alfonso Corti 12, I-20133 Milano, Italy\\
$^{25}$ AIM, CEA, CNRS, Universit\'{e} Paris-Saclay, Universit\'{e} de Paris, F-91191 Gif-sur-Yvette, France\\
$^{26}$ Instituto de F\'isica Te\'orica UAM-CSIC, Campus de Cantoblanco, E-28049 Madrid, Spain\\
$^{27}$ Institute of Space Sciences (ICE, CSIC), Campus UAB, Carrer de Can Magrans, s/n, 08193 Barcelona, Spain\\
$^{28}$ Institut d'Estudis Espacials de Catalunya (IEEC), Carrer Gran Capit\'a 2-4, 08034 Barcelona, Spain\\
$^{29}$ Dipartimento di Fisica e Scienze della Terra, Universit\'a degli Studi di Ferrara, Via Giuseppe Saragat 1, I-44122 Ferrara, Italy\\
$^{30}$ Aix-Marseille Univ, CNRS/IN2P3, CPPM, Marseille, France\\
$^{31}$ Department of Astronomy, Cornell University, Ithaca, NY, 14853, USA\\
$^{32}$ Mullard Space Science Laboratory, University College London, Holmbury St Mary, Dorking, Surrey RH5 6NT, UK\\
$^{33}$ Dipartimento di Fisica e Astronomia `Augusto Righi' - Alma Mater Studiorum Universit\'a di Bologna, Viale Berti Pichat 6/2, I-40127 Bologna, Italy\\
$^{34}$ INFN-Padova, Via Marzolo 8, I-35131 Padova, Italy\\
$^{35}$ Dipartimento di Fisica e Astronomia `G.Galilei', Universit\'a di Padova, Via Marzolo 8, I-35131 Padova, Italy\\
$^{36}$ Space Science Data Center, Italian Space Agency, via del Politecnico snc, 00133 Roma, Italy\\
$^{37}$ Astrophysics Research Institute, Liverpool John Moores University, 146 Brownlow Hill, Liverpool L3 5RF, UK\\
$^{38}$ Universit\'e St Joseph; UR EGFEM, Faculty of Sciences, Beirut, Lebanon\\
$^{39}$ School of Physics and Astronomy, Queen Mary University of London, Mile End Road, London E1 4NS, UK\\
$^{40}$ INFN-Sezione di Torino, Via P. Giuria 1, I-10125 Torino, Italy\\
$^{41}$ Dipartimento di Fisica, Universit\'a degli Studi di Torino, Via P. Giuria 1, I-10125 Torino, Italy\\
$^{42}$ INFN-Sezione di Roma, Piazzale Aldo Moro, 2 - c/o Dipartimento di Fisica, Edificio G. Marconi, I-00185 Roma, Italy\\
$^{43}$ INAF-Osservatorio Astronomico di Roma, Via Frascati 33, I-00078 Monteporzio Catone, Italy\\
$^{44}$ Jet Propulsion Laboratory, California Institute of Technology, 4800 Oak Grove Drive, Pasadena, CA, 91109, USA\\
$^{45}$ Max Planck Institute for Extraterrestrial Physics, Giessenbachstr. 1, D-85748 Garching, Germany\\
$^{46}$ Departamento de F\'isica, FCFM, Universidad de Chile, Blanco Encalada 2008, Santiago, Chile\\
$^{47}$ Institute of Cosmology and Gravitation, University of Portsmouth, Portsmouth PO1 3FX, UK\\
$^{48}$ Universit\"ats-Sternwarte M\"unchen, Fakult\"at f\"ur Physik, Ludwig-Maximilians-Universit\"at M\"unchen, Scheinerstrasse 1, 81679 M\"unchen, Germany\\
$^{49}$ INAF-Osservatorio Astrofisico di Torino, Via Osservatorio 20, I-10025 Pino Torinese (TO), Italy\\
$^{50}$ INFN-Sezione di Roma Tre, Via della Vasca Navale 84, I-00146, Roma, Italy\\
$^{51}$ Department of Mathematics and Physics, Roma Tre University, Via della Vasca Navale 84, I-00146 Rome, Italy\\
$^{52}$ INAF-Osservatorio Astronomico di Capodimonte, Via Moiariello 16, I-80131 Napoli, Italy\\
$^{53}$ Centro de Astrof\'{\i}sica da Universidade do Porto, Rua das Estrelas, 4150-762 Porto, Portugal\\
$^{54}$ Instituto de Astrof\'isica e Ci\^encias do Espa\c{c}o, Universidade do Porto, CAUP, Rua das Estrelas, PT4150-762 Porto, Portugal\\
$^{55}$ Institut de F\'{i}sica d'Altes Energies (IFAE), The Barcelona Institute of Science and Technology, Campus UAB, 08193 Bellaterra (Barcelona), Spain\\
$^{56}$ Department of Physics `E. Pancini', University Federico II, Via Cinthia 6, I-80126, Napoli, Italy\\
$^{57}$ INFN section of Naples, Via Cinthia 6, I-80126, Napoli, Italy\\
$^{58}$ Dipartimento di Fisica e Astronomia `Augusto Righi' - Alma Mater Studiorum Università di Bologna, via Piero Gobetti 93/2, I-40129 Bologna, Italy\\
$^{59}$ INAF-Osservatorio Astrofisico di Arcetri, Largo E. Fermi 5, I-50125, Firenze, Italy\\
$^{60}$ Institut national de physique nucl\'eaire et de physique des particules, 3 rue Michel-Ange, 75794 Paris C\'edex 16, France\\
$^{61}$ Centre National d'Etudes Spatiales, Toulouse, France\\
$^{62}$ Institute for Astronomy, University of Edinburgh, Royal Observatory, Blackford Hill, Edinburgh EH9 3HJ, UK\\
$^{63}$ Jodrell Bank Centre for Astrophysics, School of Physics and Astronomy, University of Manchester, Oxford Road, Manchester M13 9PL, UK\\
$^{64}$ European Space Agency/ESRIN, Largo Galileo Galilei 1, 00044 Frascati, Roma, Italy\\
$^{65}$ ESAC/ESA, Camino Bajo del Castillo, s/n., Urb. Villafranca del Castillo, 28692 Villanueva de la Ca\~nada, Madrid, Spain\\
$^{66}$ Univ Lyon, Univ Claude Bernard Lyon 1, CNRS/IN2P3, IP2I Lyon, UMR 5822, F-69622, Villeurbanne, France\\
$^{67}$ Aix-Marseille Univ, CNRS, CNES, LAM, Marseille, France\\
$^{68}$ Departamento de F\'isica, Faculdade de Ci\^encias, Universidade de Lisboa, Edif\'icio C8, Campo Grande, PT1749-016 Lisboa, Portugal\\
$^{69}$ Instituto de Astrof\'isica e Ci\^encias do Espa\c{c}o, Faculdade de Ci\^encias, Universidade de Lisboa, Campo Grande, PT-1749-016 Lisboa, Portugal\\
$^{70}$ Department of Astronomy, University of Geneva, ch. d\'Ecogia 16, CH-1290 Versoix, Switzerland\\
$^{71}$ Department of Physics, Oxford University, Keble Road, Oxford OX1 3RH, UK\\
$^{72}$ Istituto Nazionale di Astrofisica (INAF) - Osservatorio di Astrofisica e Scienza dello Spazio (OAS), Via Gobetti 93/3, I-40127 Bologna, Italy\\
$^{73}$ Istituto Nazionale di Fisica Nucleare, Sezione di Bologna, Via Irnerio 46, I-40126 Bologna, Italy\\
$^{74}$ INAF-Osservatorio Astronomico di Padova, Via dell'Osservatorio 5, I-35122 Padova, Italy\\
$^{75}$ INAF-Osservatorio Astronomico di Brera, Via Brera 28, I-20122 Milano, Italy\\
$^{76}$ Institute of Theoretical Astrophysics, University of Oslo, P.O. Box 1029 Blindern, N-0315 Oslo, Norway\\
$^{77}$ Leiden Observatory, Leiden University, Niels Bohrweg 2, 2333 CA Leiden, The Netherlands\\
$^{78}$ von Hoerner \& Sulger GmbH, Schlo{\ss}Platz 8, D-68723 Schwetzingen, Germany\\
$^{79}$ Max-Planck-Institut f\"ur Astronomie, K\"onigstuhl 17, D-69117 Heidelberg, Germany\\
$^{80}$ Institut d'Astrophysique de Paris, 98bis Boulevard Arago, F-75014, Paris, France\\
$^{81}$ Department of Physics and Helsinki Institute of Physics, Gustaf H\"allstr\"omin katu 2, 00014 University of Helsinki, Finland\\
$^{82}$ European Space Agency/ESTEC, Keplerlaan 1, 2201 AZ Noordwijk, The Netherlands\\
$^{83}$ NOVA optical infrared instrumentation group at ASTRON, Oude Hoogeveensedijk 4, 7991PD, Dwingeloo, The Netherlands\\
$^{84}$ Argelander-Institut f\"ur Astronomie, Universit\"at Bonn, Auf dem H\"ugel 71, 53121 Bonn, Germany\\
$^{85}$ Institute for Computational Cosmology, Department of Physics, Durham University, South Road, Durham, DH1 3LE, UK\\
$^{86}$ Universit\'e C\^ote d'Azur, Observatoire de la C\^ote d'Azur, CNRS, Laboratoire Lagrange, Bd de l'Observatoire, CS 34229, 06304 Nice cedex 4, France\\
$^{87}$ California institute of Technology, 1200 E California Blvd, Pasadena, CA 91125, USA\\
$^{88}$ Observatoire de Sauverny, Ecole Polytechnique F\'ed\'erale de Lau- sanne, CH-1290 Versoix, Switzerland\\
$^{89}$ Department of Physics and Astronomy, University of Aarhus, Ny Munkegade 120, DK-8000 Aarhus C, Denmark\\
$^{90}$ Perimeter Institute for Theoretical Physics, Waterloo, Ontario N2L 2Y5, Canada\\
$^{91}$ Department of Physics and Astronomy, University of Waterloo, Waterloo, Ontario N2L 3G1, Canada\\
$^{92}$ Centre for Astrophysics, University of Waterloo, Waterloo, Ontario N2L 3G1, Canada\\
$^{93}$ Institute of Space Science, Bucharest, Ro-077125, Romania\\
$^{94}$ Centro de Investigaciones Energ\'eticas, Medioambientales y Tecnol\'ogicas (CIEMAT), Avenida Complutense 40, 28040 Madrid, Spain\\
$^{95}$ Instituto de Astrof\'isica e Ci\^encias do Espa\c{c}o, Faculdade de Ci\^encias, Universidade de Lisboa, Tapada da Ajuda, PT-1349-018 Lisboa, Portugal\\
$^{96}$ Universidad Polit\'ecnica de Cartagena, Departamento de Electr\'onica y Tecnolog\'ia de Computadoras, 30202 Cartagena, Spain\\
$^{97}$ Kapteyn Astronomical Institute, University of Groningen, PO Box 800, 9700 AV Groningen, The Netherlands\\
$^{98}$ Infrared Processing and Analysis Center, California Institute of Technology, Pasadena, CA 91125, USA\\
$^{99}$ Dipartimento di Fisica - Sezione di Astronomia, Universit\'a di Trieste, Via Tiepolo 11, I-34131 Trieste, Italy\\
$^{100}$ INAF, Istituto di Radioastronomia, Via Piero Gobetti 101, I-40129 Bologna, Italy\\
$^{101}$ University of Lyon, UCB Lyon 1, CNRS/IN2P3, IUF, IP2I Lyon, France\\
$^{102}$ INAF-Istituto di Astrofisica e Planetologia Spaziali, via del Fosso del Cavaliere, 100, I-00100 Roma, Italy\\
$^{103}$ Research Program in Systems Oncology, Faculty of Medicine, University of Helsinki, Helsinki, Finland\\
$^{104}$ Department of Physics, P.O. Box 64, 00014 University of Helsinki, Finland\\
$^{105}$ Helsinki Institute of Physics, Gustaf H{\"a}llstr{\"o}min katu 2, University of Helsinki, Helsinki, Finland\\
$^{106}$ Centre de Calcul de l'IN2P3, 21 avenue Pierre de Coubertin F-69627 Villeurbanne Cedex, France\\
$^{107}$ Institut f\"ur Theoretische Physik, University of Heidelberg, Philosophenweg 16, 69120 Heidelberg, Germany\\
$^{108}$ Zentrum f\"ur Astronomie, Universit\"at Heidelberg, Philosophenweg 12, D- 69120 Heidelberg, Germany\\
$^{109}$ INFN, Sezione di Lecce, Via per Arnesano, CP-193, I-73100, Lecce, Italy\\
$^{110}$ Department of Mathematics and Physics E. De Giorgi, University of Salento, Via per Arnesano, CP-I93, I-73100, Lecce, Italy\\
$^{111}$ Institute for Computational Science, University of Zurich, Winterthurerstrasse 190, 8057 Zurich, Switzerland\\
$^{112}$ Department of Physics, P.O.Box 35 (YFL), 40014 University of Jyv\"askyl\"a, Finland\\
$^{113}$ Department of Physics and Astronomy, University College London, Gower Street, London WC1E 6BT, UK\\
$^{114}$ School of Physics and Astronomy, Cardiff University, The Parade, Cardiff, CF24 3AA, UK\\
}

\date{}

\abstract{
    \tiny The combination and cross-correlation of the upcoming \Euclid data with cosmic microwave background (CMB) measurements is a source of great expectation since it will provide the largest lever arm of epochs, ranging from recombination to structure formation across the entire past light cone. In this work, we present forecasts for the joint analysis of \Euclid and CMB data on the cosmological parameters of the standard cosmological model and some of its extensions. This work expands and complements the recently published forecasts based on \Euclid-specific probes, namely galaxy clustering, weak lensing, and their cross-correlation. With some assumptions on the specifications of current and future CMB experiments, the predicted constraints are obtained from both a standard Fisher formalism and a posterior-fitting approach based on actual CMB data. Compared to a \Euclid-only analysis, the addition of CMB data leads to a substantial impact on constraints for all cosmological parameters of the standard $\Lambda$-cold-dark-matter model, with improvements reaching up to a factor of ten. For the parameters of extended models, which include a redshift-dependent dark energy equation of state, non-zero curvature, and a phenomenological modification of gravity, improvements can be of the order of two to three, reaching higher than ten in some cases. The results highlight the crucial importance for cosmological constraints of the combination and cross-correlation of \Euclid probes with CMB data.
}


\begin{document}

\maketitle

\section{Introduction}

The apparent accelerated expansion of the Universe at recent cosmological epochs, revealed through the luminosity-distance relation of type Ia supernovae (SN, see \citealt{2019ApJ...872L..30A} and references therein), and confirmed independently by the other main cosmological probes \citep[see][ and references therein]{2020A&A...641A...6P,2020arXiv200708991E}, is one of the greatest puzzles of modern cosmology. The ensemble of possible phenomena responsible for this acceleration constitutes the classes of dark energy (DE) and modified gravity (MG) models \citep[see e.g.\ Weinberg \& White in][]{2020PTEP.2020h3C01P}, often captured under the same umbrella term of `dark energy'. Probing and unveiling the physical nature of the DE requires us to measure its effects on both the cosmological expansion and its dynamics. Both effects leave imprints at low redshift and can thus be observed through several probes, including SN, baryon acoustic oscillations \citep{2017PhT....70l..32P}, the full power spectrum of galaxy clustering \citep[GC,][]{2019BAAS...51c.508W} and weak lensing \citep[WL,][]{2020MNRAS.493.3985M}, galaxy cluster number counts \citep[][]{2016JCAP...08..005L}, and their cross-correlations with the cosmic microwave background \citep[CMB,][]{2019MNRAS.489.1950B}. The importance of understanding the nature of DE has triggered the development of large ground-based, photometric and/or spectroscopic galaxy surveys, such as the Kilo-Degree Survey\footnote{\url{kids.strw.leidenuniv.nl}} \citep{2019arXiv191014057P}, the Dark Energy Survey\footnote{\url{darkenergysurvey.org}} \citep{2018ApJS..239...18A}, the (extended) Baryon Oscillation Spectroscopic Survey\footnote{\url{sdss.org/surveys/eboss}} \citep{2017MNRAS.470.2617A,2020MNRAS.492.4189I}, the Dark Energy Spectroscopic Instrument\footnote{\url{desi.lbl.gov}} \citep{2019AJ....157..168D}, the Javalambre PAU Survey\footnote{\url{j-pas.org}}\citep{2020arXiv200701910B}, the Vera C. Rubin Observatory\footnote{\url{vro.org}} \citep[VRO, formerly the Large Synoptic Survey Telescope,][]{2018cosp...42E1651K}, and the \textit{Roman Space Telescope}\footnote{\url{roman.gsfc.nasa.gov}} \citep[formerly the Wide Field InfraRed Survey Telescope,][]{2019arXiv190205569A}.

Joint analyses combining several or all of these low-redshift probes can lead to substantial improvement in constraining the cosmological model. Although present data are already capable of constraining the values of the parameters characterising DE to an unprecedented level \citep[see e.g.][]{2018PhRvD..98d3526A}, there are still some degeneracies between these parameters and: (i) other cosmological quantities, such as the dark matter density, neutrino masses, curvature, etc.; (ii) astrophysical -- or so-called nuisance -- parameters involved in the models, such as (tracer) biases, mass calibration, etc.; and (iii) systematic effects, especially in extensions to the standard model. Such degeneracies can be broken by not only combining but also cross-correlating galaxy surveys with complementary cosmological probes.

Among these other data, the CMB provides us with a powerful and unique window on the early history of the Universe, as well as its geometry, dynamics, and content. The Atacama Cosmology Telescope\footnote{\url{act.princeton.edu}} \citep{2019MNRAS.486.5239D}, the South Pole Telescope\footnote{\url{pole.uchicago.edu}} \citep{2020ApJ...888..119B}, the PolarBear/Simons Array \citep{2020ApJ...897...55P}, and the \Planck satellite\footnote{\url{esa.int/planck}} \citep{2020A&A...641A...6P} already provide maps of the CMB temperature anisotropies with high signal-to-noise ratio overlapping with several ground-based galaxy surveys, thus allowing for joint analyses. Among potential studies, the cross-correlation of WL or GC with CMB lensing, which is (among other things) sensitive to the angular-diameter distance to the last-scattering surface, has been extensively explored \citep{2013ApJ...771L..16H,2015PhRvD..91f2001H,2016MNRAS.459...21K,2019PhRvD..99b3508B}. Another example can be found in cross-correlations with the thermal Sunyaev-Zeldovich (tSZ) signal, a secondary effect imprinted on top of the primordial CMB anisotropies, which has been measured with increasing precision as well \citep{2015JCAP...09..046M,2017MNRAS.471.1565H,2020MNRAS.492.4780O,2021arXiv210207701Y}. Additionally, several future CMB experiments, such as the Simons Observatory\footnote{\url{simonsobservatory.org}} \citep[SO,][]{Ade_2019_SimonsObs} and the so-called CMB-Stage 4 project\footnote{\url{cmb-s4.org}} \citep[CMB-S4,][]{Abazajian:2016yjj}, are planned for the next decade and will yield highly improved high-resolution measurements of the CMB intensity and polarisation. The expected improvement on the measurement of the lensing signal from the CMB photons will increase the constraining power of this probe and of its correlation with galaxy-survey-based tracers.

Within the next few years, the \Euclid satellite\footnote{\url{euclid-ec.org}} will be surveying the sky, mapping the large-scale structure (LSS) of the Universe with unprecedented precision, depth, and coverage. Over a volume corresponding to about $15\,000\,\deg^2$ on the sky and up to redshifts of $z\sim2$, the \Euclid Near Infrared Spectrometer and Photometer (\Euclid-NISP) will be able to measure up to 30 million spectroscopic redshifts \citep{Pozzetti:2016}, which can be used for GC measurements, while the \Euclid Visible Instrument (\Euclid-VIS) will measure 2 billion photometric galaxy images enabling WL observations \citep[for more details, see][]{Euclid_TWG,Euclid-r}. The unprecedented quality of the \Euclid data will offer a new insight into the late Universe, more specifically on the growth and evolution of large-scale cosmic structures and on the expansion history of the Universe. Hence, \Euclid data will enable us to make a leap forward in our understanding of the evolution of the late Universe and of the nature of the elusive DE.

Motivated by the need to assess the future performance of \Euclid and its dependence on the design of the instruments, the Euclid Consortium has dedicated considerable efforts over the years towards providing reliable and realistic forecasts for the expected accuracy of cosmological measurements. First produced in the Assessment Phase Report \citep[][colloquially known as the Yellow Book]{Euclid-y}, and later refined in the Definition Study Report \citep[][Red Book]{Euclid-r}, the \Euclid forecasts have been recently updated by \citet[][\citetalias{Euclid_IST} in the following]{Euclid_IST}, including a comprehensive comparison of different and independent forecasting codes. \citetalias{Euclid_IST} updated the specifications of the satellite with a much higher degree of precision, specified in detail the assumptions used in calculations, and explored relevant cosmological models, thus providing a complete picture of the \Euclid capabilities with respect to the late Universe observables, namely GC, WL, and their cross-correlations. Furthermore, \citetalias{Euclid_IST} provided a robust set of Fisher-matrix predictions, greatly improving upon previous work in terms of precision and sophistication.

In the future, not only will we jointly use WL and GC for the cosmological analysis of \Euclid data, but we will naturally make use of the CMB information available at that time. As shown in studies performed with CMB and ground-based survey data, the combination of the two datasets provides a great lever arm in time to constrain cosmological models. Moreover, the cross-correlations between the late-time LSS and CMB observables, arising mostly from the secondary anisotropies of CMB photons (see \citealt{2008RPPh...71f6902A} for a review), provide additional cosmological information at later times. Exploiting the cross-correlation between \Euclid and the CMB data will additionally help to reduce the impact of potential systematic effects in the datasets. The exploration and preparation for this joint analysis is the raison d'être of the `CMB cross-correlations' Science Working Group of the Euclid Consortium, responsible for the inception and the realisation of the present work.

Here, we extend the work done by \citetalias{Euclid_IST} and forecast the expected precision on cosmological parameters achievable from the complete combination and cross-correlation between \Euclid and CMB data. Concerning the main \Euclid probes, we adopt the relevant recipes and assumptions for \Euclid GC (both spectroscopic and photometric, thereafter dubbed GCs and GCp, respectively), WL and their cross-correlations. We extend those to include CMB fields, namely total intensity, polarisation and lensing, as well as all possible \ExC observables, based on the same updated specifications for the \Euclid survey.

This work is organised as follows. In Sect.~\ref{sec:Cosmology}, we describe the cosmological models considered in our forecasts and the parameters that characterise them. In Sect.~\ref{sec:Fisher}, we introduce the Fisher matrix formalism (as well as an alternative posterior-fitting method) in order to estimate expected uncertainties on cosmological parameters, and we describe our methodology for the computation of all observables relevant to the full \ExC combination. In Sect.~\ref{sec:numtools}, we present the forecasting codes included in our analysis and describe in detail the code comparison procedure that we performed, following the same guidelines as in \citetalias{Euclid_IST}. In Sect.~\ref{sec:results}, we present the final cosmological parameter forecasts for the \Euclid and CMB probes, considering first the cross-correlation with the lensing signal from the CMB separately, since it is a natural counterpart to the \Euclid probes of the LSS, and then the full joint analysis of \Euclid and CMB probes. Finally, we present our conclusions in Sect.~\ref{sec:conclusions}.


\section{Cosmological context}\label{sec:Cosmology}

This section aims to describe the different cosmological models explored in these forecasts, and their associated free parameters. The present work represents an extension and completion of the \Euclid-only study described in \citetalias{Euclid_IST}, hence for consistency we follow the same conventions for parameters and instrument specifications and we investigate the same cosmological models. To avoid unnecessary repetitions with \citetalias{Euclid_IST}, we only report here some essential points for convenience, and refer the interested reader to section 2 of \citetalias{Euclid_IST} for additional details.

The spatially flat $\Lambda$CDM model is the baseline case considered in this work; it can be described by a minimal set of six parameters. For consistency with \citetalias{Euclid_IST}, our choice of free parameters includes: (i) the dimensionless Hubble parameter $h$, defined as $H_0/(100\,\kmsMpc)$ where $H_0$ is the Hubble rate at the present time; (ii) the total matter density parameter at present time, $\Omega_{\rm m, 0}$, defined as the current total matter density divided by the critical density $\rho_{\rm c} = 3H_0^2/(8\pi G)$; (iii) the same density parameter as above but for baryons, $\Omega_{\rm b, 0}$; (iv) the spectral index of the primordial power spectrum of scalar perturbations, denoted by $n_{\rm s}$; (v) the amplitude of matter density fluctuations at the present time through the $\sigma_8$ parameter\footnote{We note that this parameter is often dropped in favour of $A_{\rm s}$ when considering CMB observables}; (vi) the optical depth of reionisation $\tau$, to which \Euclid probes are insensitive (and thus was not considered in \citetalias{Euclid_IST}), but which is crucial for CMB studies.

We include in this baseline model a so-called `minimal massive neutrino' scenario \citep[used as a baseline notably in][]{2016A&A...594A..13P}, with a single massive neutrino with 0.06 eV mass and two massless neutrinos. In our analysis, this neutrino mass is never considered as a free parameter.\footnote{\citetalias{Euclid_IST} verified that another choice for the neutrino mass (within a reasonable range permitted by current data) leads to nearly no change in cosmological forecasts.} In addition to the aforementioned baseline parameters, we consider the following three extensions to the flat-$\Lambda$CDM model, each accompanied with its new parameter(s). First, we relax the assumption of spatial flatness by allowing the curvature of the Universe to be non-zero. This is equivalent to varying $\Omega_{\rm DE, 0}$, the DE density parameter, while both $\Omega_{\rm m, 0}$ and $h$ are kept constant (see \citetalias{Euclid_IST} for details). Therefore, we use $\Omega_{\rm DE, 0}$ as a new parameter in the relevant sections of our analysis. Second, we consider the possibility of DE being dynamical in time, with a redshift-dependent equation of state $w_{\de}(z)$ following the so-called CPL parametrisation \citep{2001IJMPD..10..213C,2003PhRvL..90i1301L}, which is $w_{\de}(z) = w_0 + w_a {z}/{(1+z)}$. Two new free parameters are thus introduced, with $w_0$ being the present ($z=0$) value of the equation of state, and $w_a$ a measure of its time variation. Finally, we allow a deviation from the standard growth of structure via a departure of the so-called growth-index parameter $\gamma$ \citep{Lahav:1991wc, 2005PhRvD..72d3529L} from its standard value of $\gamma = 6/11$. We use this parametrisation to ease the comparison with the \Euclid forecasts in the Red Book and for compatibility with \citetalias{Euclid_IST}. However, although widely used, the $\gamma$ parametrisation provides an incomplete and simplified description of the evolution of perturbations (which in general requires at least two degrees of freedom as a function of time and space), as also pointed out in \citetalias{Euclid_IST}. It is valid only on sub-horizon scales and therefore not optimal for the CMB, whose observational window extends to super-horizon scales. In practice, we implemented it inside the Boltzmann codes used in our analysis, following the approach of \citet{2011JCAP...08..005H}.

We summarise in Table~\ref{tab:cosmofid} the fiducial values of all aforementioned parameters that we use throughout the present work.
\begin{table*}[h!]
    \caption{Fiducial values of the cosmological parameters considered.}
    \centering
    {\renewcommand{\arraystretch}{1.2}
    \begin{tabular}{ccccccccccccc}
    \hline
    \multicolumn{7}{c}{Baseline} & & \multicolumn{4}{c}{Extensions}\\
    \cline{1-7} \cline{9-12}
     $\Omega_{\rm b, 0}$ & $\Omega_{\rm m, 0}$ & $h$ & $n_{\rm s}$ & $\sigma_{8}$ & $\tau$ & $\sum{m_\nu}$ [eV] & & $\Omega_{\rm DE,0}$ & $w_0$ & $w_a$ & $\gamma$ \\ 
    ($\omega_{\rm b, 0}$) & ($\omega_{\rm m, 0}$) & & & & & (fixed) & & & & & \\
    \hline
    \hline
     $0.05$ & $0.32$ & $0.67$ & $0.96$ & $0.816$ & $0.058$ & $0.06$ & & $0.68$ & $-1$ & $0$ & $6/11$ \\
    ($0.022445$) & ($0.143648$) & & & & & & & & & & \\
    \hline
    \end{tabular}
    }
    \tablefoot{We show here the parameter values of our fiducial cosmological model, both in the baseline-$\Lambda$CDM case and in the extensions that we consider. Values are chosen to be identical to the ones in \citetalias{Euclid_IST}. As mentioned in the text, it should be noted that for non-flat cosmological models, $\Omega_{\rm DE,0}$ is also varied in conjunction with $\Omega_{\rm K, 0}$ (so as to keep all other parameters constant).}
    \label{tab:cosmofid}
\end{table*}


\section{Forecasting formalism}\label{sec:Fisher}

In this section, we give an overview of our forecasting formalism and define specific quantities that are used throughout this work. We also present the detailed recipes we adopted during our implementation of the Fisher matrix formalism, and for the computation of forecasts of the different cosmological probes considered in the joint \ExC analysis.

\subsection{General Fisher formalism} \label{sec:Fisher:general}

We are interested in quantifying the ability of \Euclid and CMB data to constrain the parameters of our minimal $\Lambda$CDM cosmological model and several of its extensions (see Sect.~\ref{sec:Cosmology}). In order to do so, we use a standard Fisher matrix approach \mbox{\citep{1995PhDT........19B,1996ApJ...465...34V,TTH97}} where the elements of the Fisher matrix $\tenss{F}$ are defined as (minus) the expectation value of the second derivative of the natural logarithm of the likelihood $\mathcal{L}$ with respect to the model parameters of interest (denoted $\theta_i$ here), evaluated at their fiducial values $\theta_{i, \rm fid}$:
\begin{equation}\label{eq:FMorig}
    \tenss{F}_{\alpha\beta}=-\left.\left\langle \frac{\partial^2 \ln \mathcal{L}}{\partial\theta_{\alpha}\partial \theta_{\beta}} \right\rangle\right|_{\theta_i=\theta_{i, \rm fid}} \,.
\end{equation}
The inverse of the resulting Fisher matrix is then a good approximation of the covariance of the posterior distribution of model parameters (that one would have obtained via for example standard Markov chain Monte Carlo techniques, hereafter MCMC), ensemble-averaged over many realisations of the data. If the probability distribution of the observables used to build the likelihood is Gaussian, then the Fisher matrix can be computed analytically:
\begin{equation}\label{eq:realFM}
    \tenss{F}_{\alpha\beta}=\frac{1}{2}\,\text{Tr}\left[\tenss{C}^{-1}\frac{\partial\tenss{C}}{\partial
    \theta_{\alpha}}\tenss{C}^{-1}\frac{\partial\tenss{C}}{\partial
    \theta_{\beta}}\right]+\frac{\partial
    \vec{\mu}^{\sf T}}{\partial \theta_{\alpha}}\tenss{C}^{-1}\frac{\partial
    \vec{\mu}}{\partial \theta_{\beta}} \,,
\end{equation}
where Tr stands for the trace operator for a (square) matrix, $\vec{\mu}$ is the vector of expectation values for the observables considered, $\tenss{C}$ is their theoretical covariance, and all derivatives are evaluated at the fiducial point $\theta_{i, \rm fid}$. In the present study, our observables are the estimates -- denoted by $\mathcal{\hat{C}_\ell^{XY}}$ -- of all possible angular auto- and cross-power spectra with observables $\mathcal{X}$ and $\mathcal{Y}$ drawn from the following list: CMB temperature; CMB polarisation (\textit{E} modes only); CMB lensing; GCp\footnote{GCs is considered as an independent probe, and its Fisher matrix is directly added to our final computation using the publicly available results of \citetalias{Euclid_IST}. See also the last paragraph of Sect.~\ref{sec:euclid_obs}.}; and WL. Consequently, the $\vec{\mu}$ vector is a concatenation of the theoretical expectations $\mathcal{C_\ell^{XY}}\equiv\langle\mathcal{\hat{C}_\ell^{XY}}\rangle$ of these estimated angular spectra, and the $\tenss{C}$ matrix corresponds to their covariance (cf. Eq.~\ref{eq:covfourth}), which includes the expected sources of error in the context of the various experiments we considered (see the next section for more details about the computation of those quantities). We note that, in practice, we keep only the second term of Eq.~(\ref{eq:realFM}) when computing our Fisher matrices. This is due to the fact that our observables, $\mathcal{\hat{C}}^\mathcal{XY}_\ell$, do not follow a Gaussian distribution but a Wishart one; properly deriving the Fisher formula for such a distribution (starting from Eq.~\ref{eq:FMorig}) leads to a formula identical to Eq.~(\ref{eq:realFM}) but without the leftmost term \citep[see e.g.][]{2013A&A...551A..88C,2020arXiv200510384B}.

Once the Fisher matrix is computed, an estimate of the covariance matrix $\tenss{M}$ of our model parameters\footnote{More specifically, by virtue of the Cramér-Rao inequality, we obtain a lower bound on the covariance of our parameters.} is given by
\begin{equation}
    \tenss{M}_{\alpha\beta}=(\tenss{F}^{-1})_{\alpha\beta} \,,
\end{equation}
and the square root of its diagonal elements yields the $1\sigma$ marginalised uncertainties on each parameter:
\begin{equation}
\sigma_{\alpha}=\sqrt{\tenss{M}_{\alpha\alpha}} \,.
\end{equation}

We note that in order to combine the \Euclid Fisher matrices with current CMB constraints (mainly \Planck in our case), the computation of theoretical CMB power spectra is required. These spectra can be made to reproduce as much as possible the characteristics (noise level, beams, etc.) of the actual existing CMB data, and should be computed using the same fiducial model as the one used for the \Euclid Fisher matrices. However, one may argue that this method may not accurately reproduce all the nuances of the actual CMB data. As a matter of fact, best-fit values of cosmological parameters for the \Planck data have already been determined, and slightly differ (depending on the exact dataset used) from the ones chosen by \citetalias{Euclid_IST} to build the \Euclid Fisher matrices. Moreover, the actual full \Planck likelihood is available and has been thoroughly sampled via MCMC methods. We discuss in Sect.~\ref{sec:fittedFish} an alternative to traditional Fisher forecasting, making use of real CMB data. We stress, however, that this method is not be applicable when considering forecasts involving future CMB surveys.

\subsection{Recipe for the \Euclid observables}

\label{sec:euclid_obs}

As mentioned in Sect.~\ref{sec:Fisher:general}, building our Fisher matrix implies the computation of the theoretical expectation values of the set of observables (for our fiducial models) as well as their expected covariances and derivatives with respect to the considered model parameters. Regarding the main \Euclid cosmological probes -- namely photometric and spectroscopic galaxy clustering (GCp and GCs), WL, and the cross-correlation between GCp and WL -- recipes for their computation are explained in great detail in section~3 of \citetalias{Euclid_IST}. We summarise here the main points, and focus on the specific aspects relevant to our endeavour.

Thanks to the unprecedented precision and depth of the future \Euclid survey, the WL and GC signals are expected to be measured well over a series of redshift bins; in our analysis, we consider five and ten photometric bins, respectively, for the pessimistic and optimistic Euclid scenarios. The theoretical modelling of the WL, GCp and cross-correlation power spectra in each of those redshift bins consists of computing integrals of the following form, assuming the flat-sky and Limber approximations:
\begin{equation}
\begin{split}
    \mathcal{C}^\mathcal{XY}_\ell&=c\int{\frac{\mathrm d z}{H(z)r^2(z)} \,
     \mathcal{W}^\mathcal{X}(z) \, \mathcal{W}^\mathcal{Y}(z) \, P_{\delta\delta}\left(\frac{\ell+1/2}{r(z)},z\right)}\\
     &+\mathcal{N}^\mathcal{XY}_\ell \,.
\end{split}
 \label{eq:cl}
\end{equation}
In the above expression, the letters $\mathcal{X}$ and $\mathcal{Y}$ can stand either for WL$_i$ or GCp$_i$ (the subscript $i$ referring to the $i$th redshift bin considered) and $\mathcal{W}^{\mathcal{X}}$ represents the so-called `kernel' associated with observable $\mathcal{X}$. For WL, this kernel includes contributions from both the cosmic shear signal ($\gamma$) and the intrinsic alignment (${\rm IA}$) systematic effect, and can be written as
\begin{equation}
    \mathcal{W}^{{\rm WL}_i}(z) = \mathcal{W}^{\gamma_{i}}(z) - \frac{\mathcal{P}_{\rm IA}\Omega_{\rm m,0}}{D(z)}\mathcal{W}^{\rm IA_{i}}(z) \,,
\end{equation}
with
\begin{align}
    &\mathcal{W}^{\gamma_{i}}(z) = \frac{3}{2}\,\frac{H_0^2}{c^2}\,\Omega_{\rm m,0}\,(1 + z)\,r(z)\int_{z}^{\infty} \mathrm dz^{\prime}\,\frac{n_i (z^{\prime})}{\bar{n}_i}\left[ 1 - \frac{r(z)}{r(z^{\prime})} \right] \,,\\
    &\mathcal{W}^{\rm IA_{i}}(z) = \frac{n_i (z)}{\bar{n}_i}\,\frac{H(z)}{c} \,.
    \label{eq:w_wl}
\end{align}
The kernel corresponding to the GCp is
\begin{equation}
    \mathcal{W}^{{\rm GCp}_i} (z)=b_i(z)\,\frac{n_i (z)}{\bar{n}_i}\,\frac{H(z)}{c} \,.
    \label{eq:gcp_kernel}
\end{equation}

In these expressions, $n_i(z)$ stands for the observed number density of galaxies in the $i$th redshift bin. It is given by the convolution of the true galaxy distribution $n(z)$ with the photometric redshift error, the latter being characterised by the probability $p_{\rm ph}(z_{\rm p}|z)$ of a galaxy at a (true) redshift $z$ to be measured via photometry at a redshift $z_{\rm p}$. Furthermore, ${\bar{n}_i}$ is the galaxy surface density normalising $n_i(z)$ and $D(z)$ is the linear growth factor. $\mathcal{P}_{\rm IA}$ encapsulates a specific intrinsic alignment model described in \citetalias{Euclid_IST}, and introduces three nuisance parameters named $\mathcal{A}_{\rm IA}$, $\eta_{\rm IA}$ and $\beta_{\rm IA}$ in the Fisher analyses. The specific forms of all the aforementioned functions are given in \citetalias{Euclid_IST}, and we note that all parameters of the redshift distribution and photometric error models are fixed to their fiducial value in our analysis. The galaxy bias $b_i(z)$ is assumed to be scale-independent and constant in any given redshift bin, with fiducial values $b_i=\sqrt{1+z_{{\rm c},i}}$, where $z_{{\rm c},i}$ is the central redshift of the $i$th bin. The resulting 10 bias parameters $b_i$ are part of the present Fisher analysis, where they are considered as nuisance parameters. The function $H(z)$ is the usual Hubble rate as a function of redshift $z$. The total matter power spectrum $P_{\delta \delta}$ is evaluated at $k=(\ell + 1/2)/r(z)$ in accordance with the Limber approximation, where $r(z)$ is the comoving distance as a function of redshift:
\begin{equation}
    r(z)= c \int_0^z \frac{{\rm d}z'}{H(z)} \,.
    \label{eq:gcp_kernel}
\end{equation}
Measurements by \Euclid will extend down to small scales where clustering enters the non-linear regime. As a consequence, the non-linear corrections of \texttt{Halofit} \citep{2012ApJ...761..152T}, as well as the neutrino corrections of \citet{Bird2012}, are included in $P_{\delta \delta}$. Finally, $\mathcal{N}^\mathcal{XY}_{\ell}$ contains the shot-noise term:
\begin{equation}
    \mathcal{N}^\mathcal{XY}_\ell= \frac{\sigma^2}{\bar{n}_i} \,,
\end{equation}
where $\sigma=\sigma_\epsilon$ (the total intrinsic ellipticity dispersion) when $\mathcal{X}=\mathcal{Y}=$ WL$_i$, $\sigma=1$ when $\mathcal{X}=\mathcal{Y}=$ GCp$_i$, and 0 otherwise. The shot-noise amplitude is fixed throughout the analysis (and thus is not part of the nuisance parameters).

We should note that, in practice, some of the numerical codes used in the present analysis do not directly use the expression shown in Eq.~(\ref{eq:cl}) -- as detailed in Sect.~\ref{sec:numtools}. Indeed, since CMB observables are also required in our analysis, and provided that the Limber approximation cannot be applied to the computation of such observables, the following expression was used instead:
\begin{equation}
    \begin{split}
        \mathcal{C}^\mathcal{XY}_\ell &=
         4 \pi
        \int_{0}^{\infty} {\rm d}r_1\,\mathcal{W}^\mathcal{X}(r_1)
        \int_{0}^{\infty} {\rm d}r_2\,\mathcal{W}^\mathcal{Y}(r_2) \\
        &\times \int_{0}^{\infty} \frac{{\rm d}k}{k}\,\,
        \mathcal{P_R}(k)\,\,
        T_\mathcal{X}(k,r_1)\,\,j_\ell(kr_1)\,\,
        T_\mathcal{Y}(k,r_2)\,\,j_\ell(kr_2) \,.
    \end{split}
    \label{eq:realcl}
\end{equation}
This formula involves the use of more general scale- and time-dependent `transfer functions' $T(k, r)$: they characterise the time evolution (at any given scale $k$) of a perturbation of the cosmological quantity sourcing the considered observable, for example in the case of GCp the perturbation of the matter density field. Those functions are then multiplied by the primordial curvature power spectrum ($\mathcal{P_R}$) instead of the total matter power spectrum $P_{\delta \delta}$. These transfer functions usually do not have a closed-form expression and require the use of Boltzmann codes to be computed. The presence of additional integrals involving spherical Bessel functions ($j_\ell$) renders this expression much more computationally expensive. However, it holds for both \Euclid and CMB observables, as well as their inter- and intra-cross-correlations.

\begin{figure}
    \centering
    \includegraphics[width=\columnwidth]{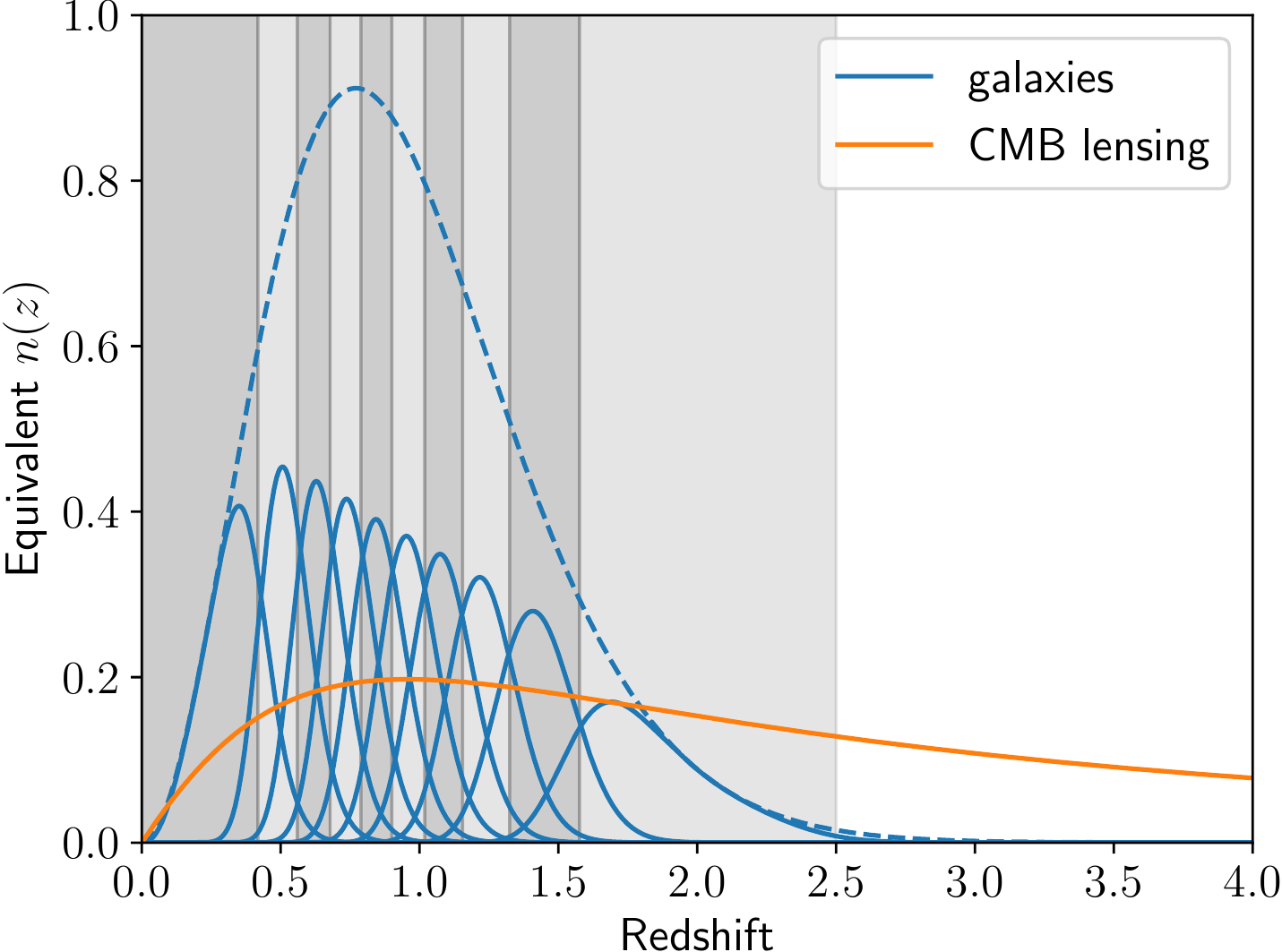}
    \caption{Redshift distribution (normalised to unit area) assumed for the galaxies of the Euclid photometric sample in dashed blue; the grey bands indicate the ten (observed) redshift bins used in the analysis, and the solid blue curves show the ten corresponding true underlying redshift distributions. The solid orange curve shows the equivalent redshift distribution (also normalised to unit area) derived from the CMB lensing kernel of Eq.~(\ref{eq:cmb_lensing_kernel}) for our fiducial cosmological model.}
    \label{fig:n_z}
\end{figure}

We report in Table~\ref{tab:WL-specifications} the specifications used to construct the aforementioned WL and GCp angular power spectra. In the present analysis, we consider two scenarios for the \Euclid probes: an optimistic case, where ten equally populated redshift bins (ranging from $z=0$ to $2.5$) are exploited and the multipole range extends from $\ell_{\rm min}=10$ to 3000 and 5000 (for GCp and WL, respectively); a more pessimistic case, where only five redshift bins (up to $z=0.9$) are considered and the highest multipole is reduced to $\ell=750$ and 1500 respectively. A visual representation of the aforementioned redshift bins is shown in Fig.~\ref{fig:n_z}, while the exact binning scheme can be found in \citetalias{Euclid_IST}.
\begin{table*}
	\centering
	\caption{Specifications for the \Euclid photometric survey.}
	\label{tab:WL-specifications}
    {\renewcommand{\arraystretch}{1.2}
	\begin{tabular}{lll}
		\hline
		& Parameter & \Euclid \\
		\hline \hline
		Survey area in the sky & $A_{\rm survey}$ & $15\,000\,\deg^2$ \\
		Sky fraction & $f_{\rm sky}$ & $0.36$ \\
		Galaxy number density & $n_{\rm g}$ & $30\,\mathrm{arcmin}^{-2}$ \\
		Total intrinsic ellipticity dispersion & $\sigma_{\epsilon}$ & 0.30 \\
		Minimum (measured) redshift & $z_{\rm min}$ & 0.001 \\
		Maximum (measured) redshift & $z_{\rm max}$ & 0.9 (pessimistic), 2.5 (optimistic) \\
		Number of redshift bins & $N_z$ & 5 (pessimistic), 10 (optimistic) \\
		Minimum multipole (WL and GC) & $\ell_{\rm min}$ & 10 \\
		Maximum multipole for WL & $\ell_{\rm max}$ & 1500 (pessimistic), 5000 (optimistic) \\
		Maximum multipole for GC & $\ell_{\rm max}$ & 750 (pessimistic), 3000 (optimistic) \\
		\hline
	\end{tabular}
    }
\end{table*}

As a final note, in accordance with the approach of \citetalias{Euclid_IST}, we consider GCs as an independent probe (thus uncorrelated to all other probes considered) throughout our whole analysis. The contribution of GCs to our final Fisher matrices is therefore accounted for via the simple addition of the corresponding Fisher matrix published by the \Euclid Collaboration. The full recipe for the GCs observables is detailed in \citetalias{Euclid_IST} and consists of the full, anisotropic, and redshift-dependent galaxy power spectrum to be derived from the \textit{Euclid} spectroscopic galaxy sample. A pessimistic and optimistic version of GCs is also considered.

\subsection{Recipe for CMB observables}
\label{sec:CMB_obs}

We consider the information available in CMB data by using power spectra from temperature and \textit{E} mode polarisation anisotropies and CMB lensing. We do not include the \textit{B} mode polarisation since it does not add information for scalar perturbations.
Our choice of observables are the various angular auto- and cross-spectra computed from those signals: $\mathcal{C}_\ell^{\rm TT}$, $\mathcal{C}_\ell^{\rm TE}$, $\mathcal{C}_\ell^{\rm EE}$, $\mathcal{C}_\ell^{\phi\phi}$, $\mathcal{C}_\ell^{\rm T\phi}$, $\mathcal{C}_\ell^{\rm E\phi}$ (where T, E and $\phi$ respectively stand for the CMB temperature, \textit{E} mode polarisation and lensing signals). The computation of those power spectra is taken care of by dedicated Boltzmann codes in our analysis, namely the popular and well-tested \texttt{CLASS} and \texttt{CAMB} public codes (see Sect.~\ref{sec:numtools} for more details).

Similarly to \Euclid observables, we contemplated a variety of scenarios for the characteristics of the CMB observables considered: a \Planck-like (nearly) full-sky survey and two types of ground-based observatories based on actual future experiments, namely the SO and CMB-S4. In the latter two cases, the fraction of the sky expected to be covered will be of order 40\,\%, translating into a lower bound of $\ell\simeq40$ for the available range of multipoles for temperature and polarisation. In order to avoid neglecting the precious amount of information contained in the CMB at large scales (and their exquisite measurement by \Planck), our hypothetical two ground-based scenarios also assume that constraints from the \Planck large-scale temperature and polarisation signals are included, via the addition of the corresponding $\mathcal{C}_\ell$s from \Planck in the Fisher analysis for multipoles $\ell < 40$. We stress, however, that measurement of the CMB lensing on such large scales is not prevented (though it is noisier) by the restricted sky fraction, thanks to the process through which the signal is recovered -- namely via high-order T and E correlations on small angles, to which all scales of the lensing signal contribute and can thus be inferred. As a consequence, for ground-based experiments we consider the full range of multipoles (starting from $\ell=2$) coming solely from the chosen experiments for CMB lensing.

Regarding the specifications of CMB measurements in all three scenarios, we adopt for the $\mathcal{C}_\ell^{\rm TT}$ and $\mathcal{C}_\ell^{\rm EE}$ spectra an isotropic noise deconvolved with the instrument beam \citep{Knox:1995dq}:
\begin{equation}
\label{eq:cmbnoise}
{\cal N}^\mathcal{XX}_\ell = w_\mathcal{XX}^{-1}\,b_{\ell}^{-2} \,, \qquad b_{\ell} = \exp\left(-\frac{1}{2}\,\ell\,(\ell+1)\,\frac{\theta_{\rm FWHM}^2}{8 \ln 2}\right) \,,
\end{equation}
where $\theta_{\rm FWHM}$ is the full-width-at-half-maximum (FWHM) of the beam given in radians and $w_{\rm TT}$, $w_{\rm EE}$ are the inverse square of the detector noise levels ($\Delta{\rm T}$ and $\Delta{\rm E}$) in $\,{\muup}\mathrm{K}\,\mathrm{arcmin}$, for temperature and polarisation, respectively. The total noise for multiple frequency channels is given by their inverse noise-weighted sum.
To estimate the CMB lensing noise, we reconstruct the minimum-variance estimator for ${\cal N}_{\ell}^{\phi\phi}$ \citep{Okamoto:2003zw} by using the publicly available code \texttt{quicklens}\footnote{\url{github.com/dhanson/quicklens}} and the noise estimates ${\cal N}_{\ell}^{\rm TT}$ and ${\cal N}_{\ell}^{\rm EE}$.

\subsubsection{\Planck}
\label{sec:plk-recipe}

For simulating our \Planck-like experiment, we aim to reproduce the \Planck 2018 results for the $\Lambda$CDM model. Our target is the corresponding baseline data combination consisting of temperature, polarisation \citep{2019arXiv190712875P}, and lensing likelihoods \citep{2018arXiv180706210P} released by the \Planck Collaboration.
 Due to the complexity of the actual \Planck data and likelihood, we have tailored the noise model and parameters in order to reproduce the real data likelihood-based uncertainties \citep{2019arXiv190712875P} with the Fisher formalism. In summary, we use the sensitivity specifications of the 143-GHz channel of the HFI instrument \mbox{\citep{2016A&A...594A...1P}} with a sky fraction $f_{\rm sky}= 0.7$ and a maximum multipole $\ell_{\rm max} = 1500$ for TT, TE, EE. In order to reproduce the (systematic-dominated) optical depth uncertainty, we inflate ${\cal N}_{\ell}^{\rm EE}$ by a factor of eight for $\ell < 30$ \citep{Bermejo-Climent:2019spz}. The \Planck-like effective noise for the CMB lensing power spectrum is obtained through the specifications of the 143\,GHz and 217\,GHz channels in \citet{2016A&A...594A...1P}. The CMB lensing power spectrum ${C}_{\ell}^{\phi \phi}$ uses a conservative range, namely $8 \le \ell \le 400$ \citep{2018arXiv180706210P}.

\subsubsection{Simons Observatory}

\begin{figure}
 \centering
 \includegraphics[width=\columnwidth]{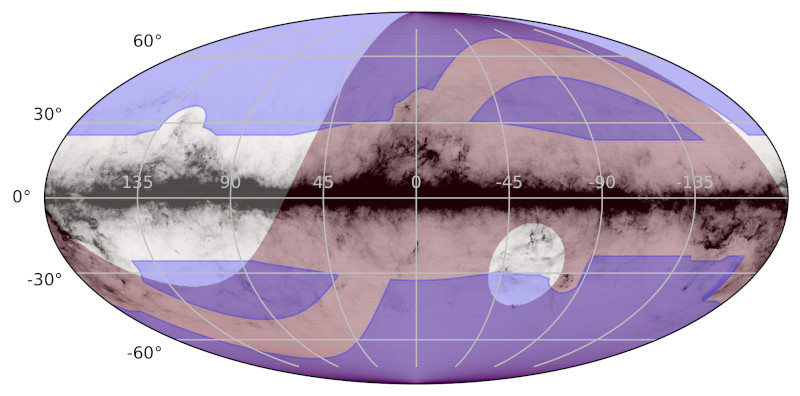}
 \caption{
    Comparison of \Euclid and SO sky coverage in Galactic coordinates. The blue, semi-transparent regions covering most of the northern and southern Galactic hemispheres away from the Galactic plane is the currently planned \Euclid sky coverage (Scaramella et al. in preparation). In addition to a Galactic cut, \Euclid also avoids regions within about $15^\circ$ of the ecliptic plane, seen here as the band running from the lower, left-hand part of the graphic to the upper, right-hand part. The region covering most of the right of the plot (with a light red hue) indicates the sky available to SO, assuming a minimum observation elevation of $40\degree$. Roughly 70\,\% of the celestial sphere is accessible to SO, though this is reduced to around 40\,\% when one makes reasonable Galactic cuts. The underlying, grey-scale map shows the Planck 545-GHz map, which gives one an indication of what regions are most contaminated by thermal emission from Galactic dust.
 }
 \label{fig:SimonsCoverage}
\end{figure}

The SO was initiated in 2016 with the goal of mapping the CMB with three 50-cm-class refracting imagers and one 6-m telescope at an altitude of 5,200\,m in the Chilean Andes. This 6-metre telescope will produce data appropriate for combination and cross-correlation with \Euclid. First-light is expected in 2022, with science operations beginning in 2023. It will ultimately cover a usable fraction of the sky of 40\,\% (cf.\ Fig.~\ref{fig:SimonsCoverage}) with beam FWHM between 0.9 and 7.4\,arcmin over the frequency range from 27 through 280\,GHz. Over the six bands spanning this frequency range, the temperature noise expectations are 71 to $54\,{\muup}\mathrm{K}\,\mathrm{arcmin}$, while the goals are a factor of $\sqrt{2}$ better \mbox{\citep{Ade_2019_SimonsObs}}. Figure~\ref{fig:SimonsCoverage} compares the sky coverage of \Euclid and SO. We use in our Fisher analysis the noise curves provided by the SO Collaboration in \citet{Ade_2019_SimonsObs}.\footnote{We use the version 3.1.0 available at \url{github.com/simonsobs/so_noise_models}.} In practice, we take the noise curves obtained with the ILC component-separation method, assuming the baseline analysis for a sky fraction of 0.4 (see Fig.~\ref{fig:SimonsCoverage}). We differ here from the formula used for \Planck, since the SO noise is modelled using the component-separation method for all channels. For our forecasts with SO, we use data over $40\le\ell\le 3000$ for TT and TE, $40\le\ell\le 5000$ for EE, and $2\le\ell\le 3000$ for $\phi \phi$ and ${\rm T}\phi$. As mentioned at the beginning of Sect.~\ref{sec:CMB_obs}, this scenario also considers the addition of large-scale data from the \Planck survey; in practice, we add this information via the first multipoles of all \textit{T}- and \textit{E}-related spectra considered, up to $\ell = 40$, with the same specifications as described earlier for the \Planck-like survey.

\subsubsection{CMB-Stage 4}

The CMB-S4 experiment, which will follow the SO (in combination with the successors to the South Pole Telescope and the current BICEP/Keck collaboration in Antarctica) is scheduled to begin taking data in 2027. We assume that it will have an additional dedicated large-aperture telescope similar to that of SO described above~\citep{Abazajian:2016yjj} and, given more stringent foreground requirements, will also cover 40\,\% of the sky, although with improved depth. We use $\Delta{\rm T} = 1\,{\muup}\mathrm{K}\,\mathrm{arcmin}$, $\Delta{\rm E}=\sqrt{2}\,{\muup}\mathrm{K}\,\mathrm{arcmin}$ and $\theta_{\rm FWHM} = 1\,{\muup}\mathrm{K}\,\mathrm{arcmin}$. CMB-S4 is expected to use data over $40\le\ell\le 3000$ in temperature and $40\le\ell\le 5000$ in polarisation. The lensing power spectrum will use data from $2\le\ell\le 3000$, and its noise curve is taken as the so-called N0 bias \citep[see e.g.][]{2017PhRvD..96f3510C}, which is computed using the \texttt{quicklens} code. Similarly to the SO case, \Planck-like large-scale information is also added in this scenario. We note that one can expect to improve upon the noise level since it is estimated by the \citet{Okamoto:2003zw} method, notably by using iterative delensing \citep[see e.g.][]{2018PhRvD..97l3540S}. Therefore, the results shown in the present work should be thought as being conservative and a lower bound on the constraints obtainable from CMB lensing.

\begin{figure}
    \centering
    \includegraphics[width=0.95\columnwidth]{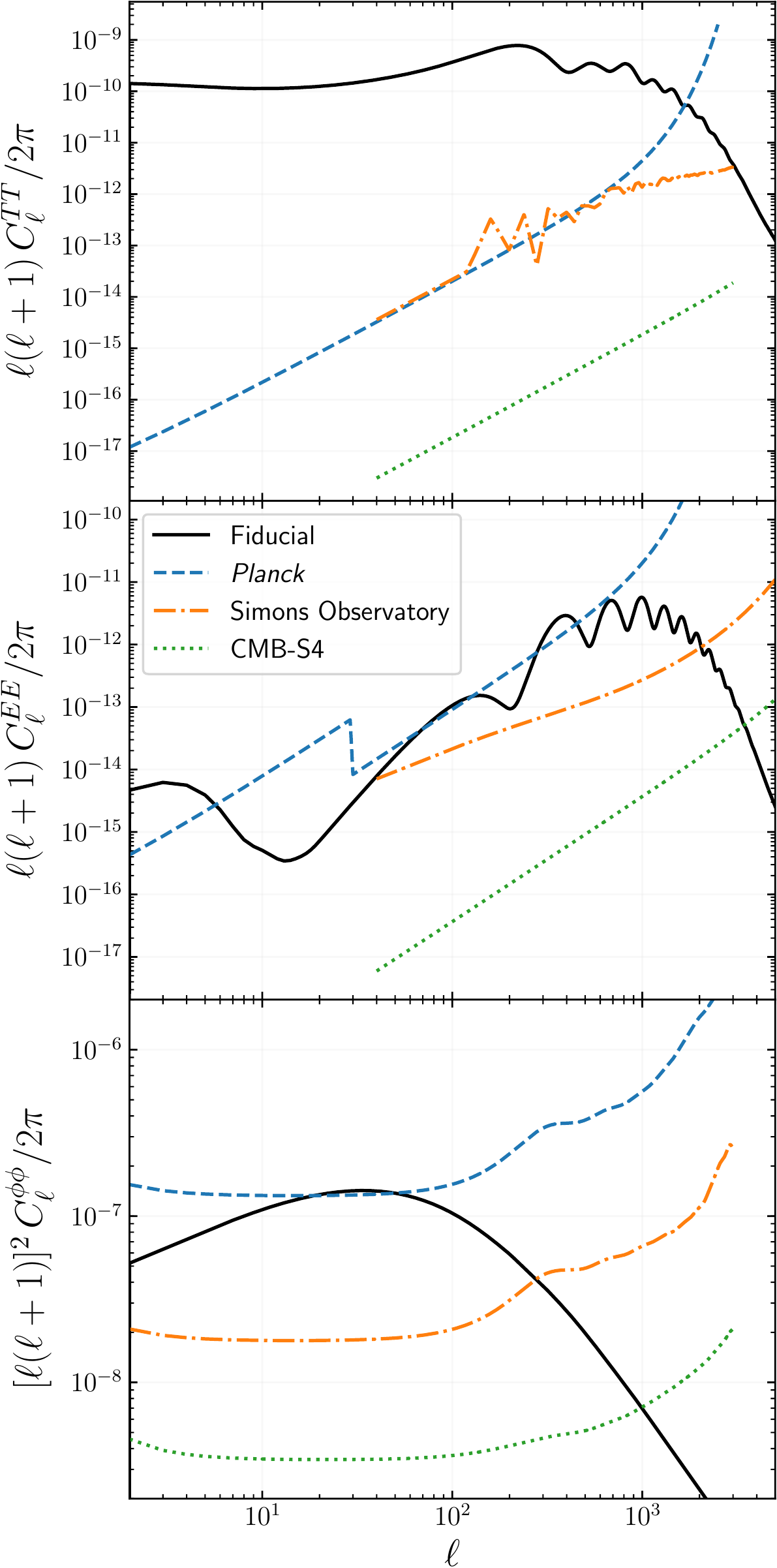}
    \caption{Noise power spectra used to reproduce the CMB experiments considered. We show the noise curves for \Planck (dashed blue lines), SO (dash-dotted orange lines), and CMB-S4 (dotted green lines), as well as our fiducial $\mathcal{C}_\ell$ for reference in black. The top, middle, and bottom panels correspond, respectively, to the temperature, polarisation (we note the inflated noise at $\ell<30$ for the \Planck-like case, cf. Sect.~\ref{sec:plk-recipe}) and CMB lensing signals. Only auto-power spectra are shown, since we neglect the noise for the cross-correlation between CMB signals.}
    \label{fig:noise-CMB}
\end{figure}

All specifications for our three considered CMB experiments are summarised in Table~\ref{tab:CMB_spec}, and the noise curves for all CMB auto-spectra are shown in Fig.~\ref{fig:noise-CMB}.

\begin{table*}[h!]
    \centering
    \caption{Specifications for CMB experiments.}
    \label{tab:CMB_spec}
    {\renewcommand{\arraystretch}{1.2}
    \begin{tabular}{lllll}
    \hline
         & Parameter & \Planck & Simons Observatory & CMB+Stage 4 \\
         &  &  & $+$ \Planck low-$\ell$ & $+$ \Planck low-$\ell$ \\
    \hline
    \hline
        Sky fraction & $f_{\rm sky}$ & $0.7$ & $0.4$ & $0.4$ \\
        Beam FWHM & $\theta_{\rm FWHM}$ & 7\,arcmin & 2\,arcmin & 1\,arcmin \\
        Temperature noise & $\Delta{\rm T} \equiv (w_{\rm TT})^{-1/2}$ & $23\,{\muup}\mathrm{K}\,\mathrm{arcmin}$ & $3\,{\muup}\mathrm{K}\,\mathrm{arcmin}$ & $1\,{\muup}\mathrm{K}\,\mathrm{arcmin}$ \\
        Polarisation noise & $\Delta{\rm E} \equiv (w_{\rm EE})^{-1/2}$ & $42\,{\muup}\mathrm{K}\,\mathrm{arcmin}$ & $3\sqrt{2}\,{\muup}\mathrm{K}\,\mathrm{arcmin}$ & $\sqrt{2}\,{\muup}\mathrm{K}\,\mathrm{arcmin}$ \\
        TT multipole range & $[\ell_{\rm TT, min}, \ell_{\rm TT, max}] $ & $[2, 1500]$ & $[2, 3000]$ & $[2, 3000]$\\
        TE multipole range & $[\ell_{\rm TE, min},\ell_{\rm TE, max}] $ & $[2, 1500]$ & $[2, 3000]$ & $[2, 3000]$\\
        EE multipole range & $[\ell_{\rm EE, min},\ell_{\rm EE, max}] $ & $[2, 1500]$ & $[2, 5000]$ & $[2, 5000]$\\
       $\phi\phi$ multipole range & $[\ell_{\phi\phi, \rm min}, \ell_{\phi\phi, \rm max} ] $ & $[8, 400]$ & $[2, 3000]$ & $[2, 3000]$\\
       T$\phi$ multipole range & $[\ell_{\rm T\phi, min}, \ell_{\rm T\phi, max} ] $ & $[8, 400]$ & $[2, 3000]$ & $[2, 3000]$\\
    \hline
    \end{tabular}
    }
\end{table*}

\subsection{Recipe for \ExC observables}

The computation of the \ExC observables (namely, all possible combinations of the GCp or WL measurements with the T, E, or $\phi$ signals) can be performed in principle by using either of the two angular spectra equations described in Sect.~\ref{sec:euclid_obs}, namely Eq.~(\ref{eq:cl}) and Eq.~(\ref{eq:realcl}). In practice, due to the non-trivial nature of their transfer functions, any combination involving T or E cannot be performed via the simplified, Limber-approximated Eq.~(\ref{eq:cl}) and thus requires the use of Eq.~(\ref{eq:realcl}) -- together with a Boltzmann code. We note, however, that due to the mostly late-time nature of the CMB lensing signal, combinations involving $\phi$ can use the simplified formula for cross-spectra, using the following scale-independent CMB lensing kernel:
\begin{equation}
    \label{eq:cmb_lensing_kernel}
    \mathcal{W}^{\phi}(z) = \frac{3}{2}\,\frac{H_0^2}{c^2}\,\Omega_{\rm m,0}\,(1 + z)\,r(z)\left[ 1 - \frac{r(z)}{r(z^\ast)} \right] \,,
\end{equation}
where $r(z^\ast)$ is the comoving distance to the surface of last scattering at redshift $z^\ast$ \citep[see e.g.][]{LewisChallinor2006}. For comparison purposes, one can derive from the CMB lensing kernel an equivalent of the redshift distribution present in the GC kernel in Eq.~(\ref{eq:gcp_kernel}):
\begin{equation}
    \label{eq:cmb_lensing_kernel}
    \mathcal{W}^{\phi}(z) \equiv \frac{n_{\phi} (z)}{\bar{n}_{\phi}}\,\frac{H(z)}{c} \ \Rightarrow\ \frac{n_{\phi} (z)}{\bar{n}_{\phi}} \equiv \mathcal{W}^{\phi}(z)\,\frac{c}{H(z)} \,.
\end{equation}
We show the resulting $n_{\phi}$ in Fig.~\ref{fig:n_z} for comparison with the GC redshift distributions used in our analysis.

In total, we thus consider six cross-correlation signals between \Euclid and CMB probes, namely GCp$\times$T, GCp$\times$E, GCp$\times$$\phi$, WL$\times$T, WL$\times$E, and WL$\times$$\phi$. We stress that for all cross-signals involving WL, we properly account for the fact that the WL kernel is composed of a shear ($\gamma$) part and an IA part (cf. Sect.~\ref{sec:euclid_obs}) and that each correlates differently to CMB observables.

\subsection{Fisher-matrix implementation}
\label{sec:fisher_implementation}

A key component required for the computation of Fisher matrices is the covariance matrix of the considered observables. We restrict ourselves to a Gaussian formulation of the covariance between estimated angular power spectra, which is given by
\begin{equation}
    \begin{split}
    {\rm Cov}&\left[\mathcal{\hat{C}}^\mathcal{XY}_\ell,\mathcal{\hat{C}}^\mathcal{X^\prime Y^\prime}_{\ell^\prime}\right] = \frac{\delta^{\rm K}_{\ell\ell^\prime}}{(2 \ell + 1) f_{\rm sky}} \\
    &\times \left\{ \left[\mathcal{C}^\mathcal{XX^\prime}_\ell +
     \mathcal{N}^\mathcal{XX^\prime}_\ell\right]\left[\mathcal{C}^\mathcal{YY^\prime}_{\ell^\prime} + \mathcal{N}^\mathcal{YY^\prime}_{\ell^\prime}\right] \right. \\
    &\ \ + \left. \left[\mathcal{C}^\mathcal{XY^\prime}_\ell + \mathcal{N}^\mathcal{XY^\prime}_\ell\right]\left[\mathcal{C}^\mathcal{YX^\prime}_{\ell^\prime} + \mathcal{N}^\mathcal{YX^\prime}_{\ell^\prime}\right] \right\} \,,
    \end{split}
    \label{eq:covfourth}
\end{equation}
where $\delta^{\rm K}$ is the Kronecker delta, meaning that we assume no correlations between different multipoles. The indices $\mathcal{X}$, $\mathcal{Y}$, $\mathcal{X^\prime}$ and $\mathcal{Y^\prime}$ can be any of the observables considered, namely T, E, $\phi$, WL$_i$ or GCp$_i$. The noise terms for auto-spectra are described in Sects.~\ref{sec:euclid_obs} and \ref{sec:CMB_obs}, while all cross-spectra are assumed to have zero noise; this assumption holds as long as no given systematic effect contributes to the corresponding combination of observables. We make a conservative choice for the value of the sky fraction $f_{\rm sky}$ in Eq.~(\ref{eq:covfourth}) by setting it each time to the smallest fraction among the four $\mathcal{X}$, $\mathcal{Y}$, $\mathcal{X^\prime}$ and $\mathcal{Y^\prime}$ observables.

Given those assumptions, we can rewrite the Fisher matrix expression in Eq.~(\ref{eq:realFM}) as a sum of independent Fisher matrices, one per multipole:
\begin{equation}\label{eq:FMsum}
\tens{F}_{\alpha\beta}= \sum_{\ell} \tens{F}_{\alpha\beta,\ell} \,,
\end{equation}
where for a given multipole $\ell$, the `partial' Fisher matrix $\tens{F}_{\alpha\beta,\ell}$ is given by (see details in Sec.~\ref{sec:Fisher:general})
\begin{equation}\label{eq:FMpart}
    \tens{F}_{\alpha\beta,\ell} = \frac{\partial
    \vec{\mu}^{\sf T}}{\partial \theta_{\alpha}}\tens{C}^{-1}\frac{\partial
      \vec{\mu}}{\partial \theta_{\beta}} \,.
    \end{equation}
In this expression, the vector $\vec{\mu}$ contains all possible auto- and cross-power spectra $\mathcal{C}^\mathcal{XY}_\ell$ derivable from the considered probes for that particular multipole $\ell$, while $\tenss{C}$ is the covariance matrix of the estimated $\mathcal{\hat{C}}^\mathcal{XY}_\ell$ computed via Eq.~(\ref{eq:covfourth}).

\subsection{Fitted-Fisher approach}\label{sec:fittedFish}

As a complement to (and a form of validation of) the traditional Fisher-matrix methodology, we also adopted an alternative technique, referred to as the `fitted-Fisher approach'. It is designed to combine a theoretical Fisher matrix, such as the one computed for \Euclid, with likelihood constraints from actual data -- in our case, CMB data from the \Planck satellite. A detailed description of the method, as well as the associated results and discussion, can be found in Appendix~\ref{sec:Appendix}. Overall we find a good agreement between the use of the actual Planck data and the standard Fisher forecast described in Sections~\ref{sec:Fisher:general} to~\ref{sec:fisher_implementation}


\section{Numerical tools}\label{sec:numtools}

This section presents the description and validation of the Fisher matrix codes used in this work, each of which provides an implementation of the formalism described in Sect.~\ref{sec:Fisher}, assuming the various cosmological models given in Sect.~\ref{sec:Cosmology}. The codes are capable of computing two of the \Euclid main probes (GCp and WL) as well as CMB observables (temperature, polarisation, and lensing), and have been specifically optimised for computing the cross-correlations between the two. They follow the \Euclid and CMB survey specifications outlined earlier.

\subsection{Code descriptions}\label{sec:allcodes}

As a starting point, we provide in the following a brief description of each code, highlighting their main features. Several programming languages and approaches have been adopted, thus providing a way to test the robustness of results against implementation choices. The codes are interfaced with different Boltzmann solvers widely used by the cosmological community, namely either \texttt{CAMB sources}\footnote{\url{camb.info/sources}} or \texttt{CLASS}\footnote{\url{class-code.net}} \citep{2011JCAP...07..034B}, which adds another layer of validation to our results. \texttt{CosmicFish}\footnote{\url{cosmicfish.github.io}} \citep{2016arXiv160606268R,2016arXiv160606273R} and \texttt{CosmoSIS}\footnote{\url{bitbucket.org/joezuntz/cosmosis}} \citep{2015A&C....12...45Z} are two of the codes that were employed in \citetalias{Euclid_IST}, validated against a number of additional codes in the context of computing forecasts for the main \Euclid probes. As a consequence, those codes have not been designed to compute CMB and \ExC observables. Nonetheless, they were used in the present work as references for the computation of \Euclid observables. Full details of the two codes can be found in \citetalias{Euclid_IST}.

\texttt{BolFish}\footnote{Contact: J. R. Bermejo-Climent} is a Fisher matrix code for CMB and LSS now written in Python 3 \citep{2021PhRvD.103j3502B} and used in its previous implementations in \cite{Ballardini:2016hpi,Ballardini:2017xnt}. In its basic settings, it interfaces with a modified version of \texttt{CAMB sources}, which allows for the introduction of survey specifications (such as the number density of galaxies or the galaxy bias function), nuisance parameters and tomography. The \texttt{BolFish} code is also able to work with angular power spectra for CMB anisotropy fields, galaxy counts and their cross-correlation, or with the matter power spectrum $P(k)$ provided as external inputs. \texttt{BolFish} allows the exploration of several extensions to the standard $\Lambda$CDM model, such as the CPL DE parametrisation or massive neutrinos. By default, numerical derivatives are computed with a three-point stencil method, where the step size for each parameter is user-defined. This Fisher code allows the use of different settings (e.g. $\ell$ range) for the different angular power spectra considered, which is useful when performing a joint 2D analysis of CMB and LSS probes.

\texttt{SFX-CLASS}\footnote{\label{fn:SI}Contact: S. Ili\'c} comprises a customised version of the \texttt{CLASS} Boltzmann code, optimised for the purpose of computing LSS and LSS$\times$CMB observables. It is broadly compatible with any LSS survey specifications and adds the computation of systematic effects such as intrinsic alignments for WL calculations. It supports all cosmological models implemented in the standard version of \texttt{CLASS}, and is theoretically expandable to all extensions of \texttt{CLASS} that have been developed in the literature. It is also interfaced with a user-friendly Fisher-matrix code (written in Python), dedicated to performing the repeated calls to the aforementioned modified \texttt{CLASS} code, which are required to obtain the derivatives of the observables under consideration. It computes these derivatives via either a standard three-point stencil method or a polynomial fit of arbitrary order over an arbitrary number of points (respectively fixed to 2 and 21 in the present work) around the chosen fiducial point.

\texttt{PyCross}\textsuperscript{\ref{fn:SI}} is a standalone Python code, interfaced with a standard version of the \texttt{CLASS} code. It takes care of the computation of the angular power spectra for LSS and LSS$\times$CMB observables itself, combining intermediate outputs from \texttt{CLASS} (such as the matter power spectrum) and LSS survey specifications, without using the Limber approximation. Its main feature is the use of the so-called \texttt{FFTLog}\footnote{\url{jila.colorado.edu/~ajsh/FFTLog}} algorithm to compute the complex integrals required (see Eq.~\ref{eq:realcl}), a very efficient alternative to the brute force approach when the integrals involved contain Bessel functions. However, since the algorithm relies on the time-space separability of the considered kernels, \texttt{PyCross} cannot compute CMB angular power spectra and instead uses directly outputs from \texttt{CLASS}. The code has no Fisher computation capabilities; it was mainly used in the present work as a means to assess to what extent the calculation of angular power spectra may be impacted by the chosen method of integration, given the large-scale nature of the LSS$\times$CMB cross-correlation (and thus its sensitivity to approximations such as the Limber one).

\texttt{TomoCelle}\footnote{Contact: C. Hern\'andez-Monteagudo} is a code written in IDL with an interface to \texttt{CAMB} for the 3D power spectrum estimation, and with a \texttt{FFTLog} Fortran routine for the auto and CMB-cross angular power spectra computation. The code supports any arbitrary selection function for the survey with any generic $b(k,z)$ galaxy bias dependence. However, it lacks a Fisher-matrix module, and its use was restricted to the comparison and validation of angular power spectra.


\subsection{Code comparison and validation}\label{sec:compcodes}

A crucial step in any Fisher forecast analysis is to verify the accuracy of the numerical codes involved at every step of the computation of the final constraints. The use of multiple distinct codes (at least two) computing the same quantities is essential in order to validate them. In this context, we aimed our comparison efforts towards three main computed quantities: (i) theoretical observables, namely the angular power spectra associated with the \Euclid, CMB, and \ExC probes, which involve the various Boltzmann codes used here in their computation; (ii) their derivatives with respect to our parameters of interest (both cosmological and nuisance types), which notably relies on some arbitrary choices in the numerical differentiation method adopted; and (iii) the computation of the final Fisher matrices themselves, mostly involving matrix algebra.

We assumed that CMB observables did not require too much scrutiny from our side. Indeed, in all the codes involved in the present work, the computation of these observables is taken care of by very well-tested Boltzmann codes (namely \texttt{CAMB} and \texttt{CLASS}), which have been the subject of dedicated validation studies in the literature \citep[see e.g.][]{2011arXiv1104.2934L,2012JCAP...04..027H}. The codes were used mostly `out of the box', with precision settings high enough to not require further validation.

Regarding \Euclid observables, the outputs of the two aforementioned IST-validated codes \texttt{CosmicFish} and \texttt{CosmoSIS} acted as benchmarks for validating the \Euclid observables computed by the four other codes introduced in the previous section. Agreement was reached at the 0.1--1\,\% level over the whole range of relevant multipoles and probes considered here. The comparison of \ExC observables was carried out between the four codes specifically developed for the present work and reached a similar level of validation, despite their non-negligible conceptual and practical differences.

Considering the very large number of angular power spectra considered as observables (namely 276) and multipole range (up to $\ell=5000$), multiplied by the numerous parameters involved in our analyses (10 cosmological and 13 nuisance) a direct comparison of the derivatives among different codes proved impracticable and mostly unnecessary. We thus decided to validate those derivatives directly at the level of the Fisher matrices. Such an approach can be further justified by the observation that a potentially large relative difference between codes in the derivative of a given spectrum may not have any influence at all on the final Fisher forecasts, especially if the covariance of the corresponding spectrum is much larger that its derivative. We were able to reach agreement on the final marginalised constraints on parameters, all within 10\,\% of each other (a criterion adopted by \citetalias{Euclid_IST}) when compared to the \Euclid-only results of \citetalias{Euclid_IST}, and when comparing the various internal codes used in our analysis.


\section{Results}\label{sec:results}

In this section, we present and discuss the results of the validated Fisher matrix codes for the cosmological parameters under study, namely
\begin{displaymath}
    \null \left\{\Omega_{\rm b,0},\,\Omega_{\rm m,0},\,h,\,n_{\rm s},\,\sigma_8, \, \, \tau, \,\,\Omega_{\rm DE,\,0},\,w_0,\,w_{a},\,\gamma\right\} \,,
    \label{eq:paramRecall}
\end{displaymath}
whose fiducial values are reported in Table~\ref{tab:cosmofid}, as well as the nuisance parameters (the 10 galaxy bias and three IA parameters). More precisely, we explore the six cosmological `cases' considered by \citetalias{Euclid_IST}, namely: (i) a flat universe with cosmological constant ($\Lambda$CDM); (ii) non-flat $\Lambda$CDM; (iii) flat with time-varying DE equation of state (${w_0}{w_a}$CDM); (iv) non-flat ${w_0}{w_a}$CDM; (v) flat with time-varying DE and phenomenological modification of gravity ($w_0w_a\gamma$CDM); and (vi) non-flat $w_0w_a\gamma$CDM. The main objective of the present work is to quantify the benefits of the combination of \Euclid probes with CMB data. We thus first show as a reference in Table~\ref{tab:resEuclid} the precision reached on all the aforementioned parameters when \Euclid-only probes are considered, namely GCs, GCp, WL, and GCp$\times$WL. We recall that all computations required to obtain those results were carried out using a dedicated suite of codes, stemming from the recipes of \citetalias{Euclid_IST} -- with the exception of GCs, which was added directly at the Fisher matrix level as an independent probe, using the official public Euclid Collaboration matrices.

We present first in Fig.~\ref{fig:all_res_addCMBlens} the improvements resulting from the combination of \Euclid probes with a subset of the CMB observables, namely the CMB lensing signal only. This subset is indeed a natural choice for combination with large-scale galaxy surveys, since it represents a counterpart in the CMB to late-Universe tracers of the matter distribution. Moreover, it is often considered a somewhat cleaner probe of matter, and it is hoped that it will alleviate some of the tracer-related systematics that plague galaxy surveys. We then present in Fig.~\ref{fig:all_res_addfullCMB} the outcome of a complete joint \ExC analysis (with all CMB probes added), showcasing the constraining power of the full combination.

\begin{table*}[h!]
    \caption{Predicted constraints on cosmological parameters from \Euclid.}
    \centering
    {\renewcommand{\arraystretch}{1.2}
    \begin{tabular}{rccccccccc}
        \hline
        Model & $\Omega_{\rm b, 0}$ & $\Omega_{\rm m, 0}$ & $n_{\rm s}$ & $h$ & $\sigma_{8}$ & $\Omega_{\rm DE,0}$ & $w_0$ & $w_a$ & $\gamma$ \\
        \hline
        \hline
        & \multicolumn{9}{c}{\underline{\Euclid pessimistic}} \\
        flat $\Lambda$CDM & 0.025 & 0.0065 & 0.0052 & 0.0036 & 0.0031 & \dots & \dots & \dots & \dots \\
        non-flat $\Lambda$CDM & 0.026 & 0.0065 & 0.0054 & 0.0042 & 0.0032 & 0.0099 & \dots & \dots & \dots \\
        flat ${w_0}{w_a}$CDM & 0.031 & 0.011 & 0.0056 & 0.0046 & 0.0045 & \dots & 0.038 & 0.14 & \dots \\
        non-flat ${w_0}{w_a}$CDM & 0.031 & 0.011 & 0.0056 & 0.0047 & 0.0047 & 0.025 & 0.039 & 0.22 & \dots \\
        flat $w_0w_a\gamma$CDM & 0.038 & 0.015 & 0.0059 & 0.0047 & 0.0050 & \dots & 0.039 & 0.14 & 0.015 \\
        non-flat $w_0w_a\gamma$CDM & 0.038 & 0.015 & 0.0059 & 0.0047 & 0.0055 & 0.025 & 0.039 & 0.23 & 0.016 \\
        \hline
        & \multicolumn{9}{c}{\underline{\Euclid optimistic}} \\
        flat $\Lambda$CDM & 0.011 & 0.0025 & 0.0015 & 0.0011 & 0.0012 & \dots & \dots & \dots & \dots \\
        non-flat $\Lambda$CDM & 0.011 & 0.0031 & 0.0018 & 0.0014 & 0.0012 & 0.0064 & \dots & \dots & \dots \\
        flat ${w_0}{w_a}$CDM & 0.013 & 0.0053 & 0.0019 & 0.0014 & 0.0019 & \dots & 0.021 & 0.073 & \dots \\
        non-flat ${w_0}{w_a}$CDM & 0.013 & 0.0053 & 0.0019 & 0.0015 & 0.0020 & 0.011 & 0.021 & 0.086 & \dots \\
        flat $w_0w_a\gamma$CDM & 0.017 & 0.0083 & 0.0022 & 0.0016 & 0.0024 & \dots & 0.021 & 0.073 & 0.0077 \\
        non-flat $w_0w_a\gamma$CDM & 0.018 & 0.0085 & 0.0022 & 0.0016 & 0.0027 & 0.011 & 0.021 & 0.092 & 0.0086
    \end{tabular}
    }
    \tablefoot{We report here the predicted constraints on cosmological parameters from the joint analysis of all \Euclid probes (GCs$+$WL$+$GCp$+$GCp$\times$WL), expressed as the ratio of marginalised 1$\sigma$ uncertainties over their corresponding fiducial values (cf. Table~\ref{tab:cosmofid}). We note that for $w_a$, whose fiducial value is 0, we directly quote the 1$\sigma$ uncertainties. Because \Euclid observables alone cannot constrain $\tau$, it is absent from this table. Values shown here are not exactly identical to (but within an acceptable range of) the same quantities reported in \citetalias{Euclid_IST}, due to small numerical differences between the codes used.}
    \label{tab:resEuclid}
\end{table*}

In the aforementioned figures, for each cosmological and nuisance parameter the colour coding reflects the percentage improvement defined as
\begin{equation}
    \left(\frac{\sigma_{\rm before}}{\sigma_{\rm after}}-1\right)\times100 \,,
\end{equation}
where $\sigma_{\rm before}$ and $\sigma_{\rm after}$ are, respectively, the 1$\sigma$ uncertainties before and after adding CMB-related constraints, while the number in each square corresponds to the factor of improvement, namely
\begin{equation}
    \frac{\sigma_{\rm before}}{\sigma_{\rm after}} \,.
\end{equation}

\subsection{\Euclid and CMB lensing combination}\label{sec:resultsEplusCMB}

\begin{figure*}
    \centering
    \includegraphics[width=\textwidth]{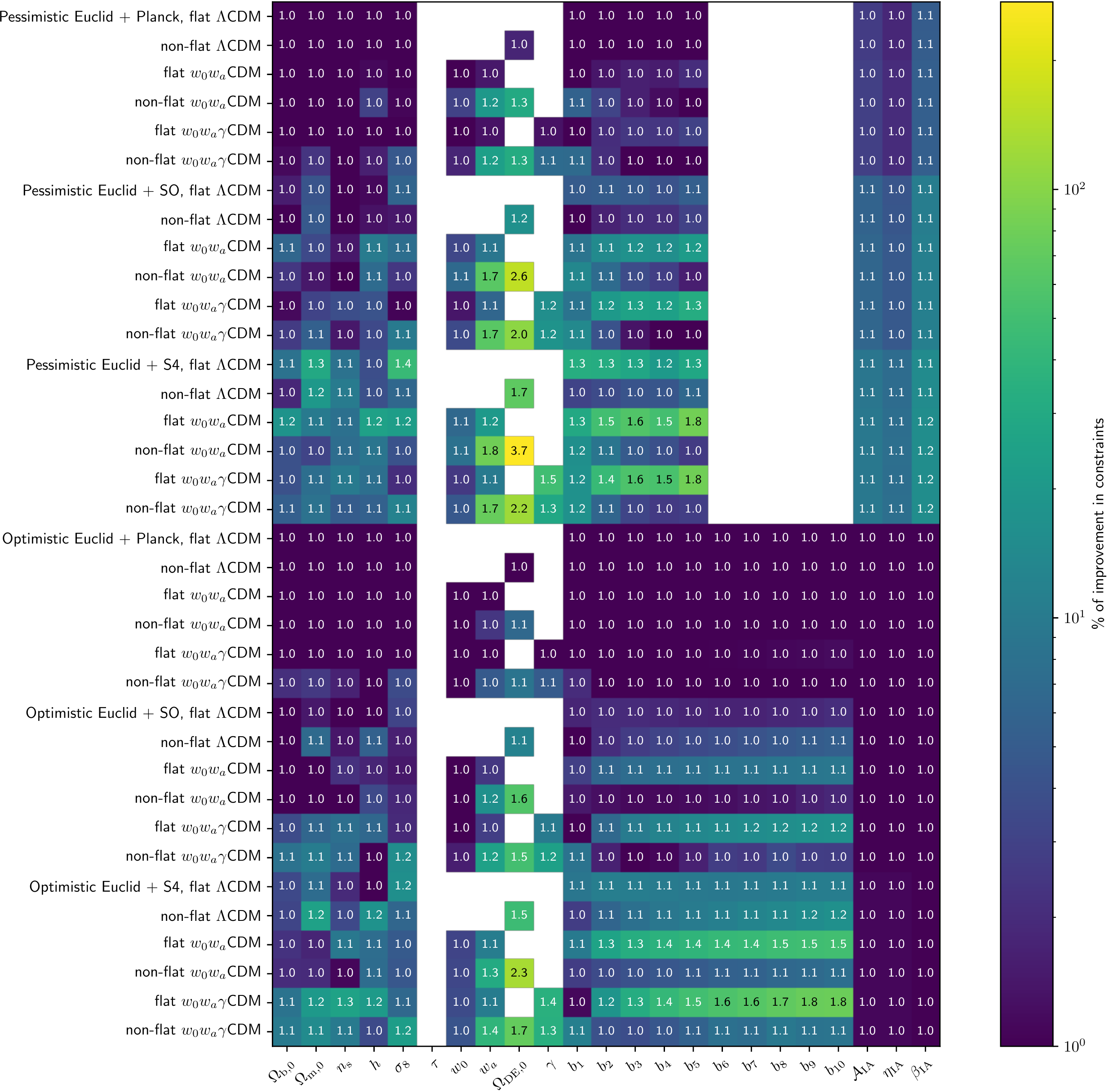}
    \caption{Predicted improvements on parameter constraints after adding CMB lensing data to \Euclid, including all cross-correlations, for all scenarios (cosmological models and survey specifications) considered in the present study. Colours indicate the percentage (cropped below 1\,\%) of improvement in constraints, while numbers show the factors in the uncertainties reduction (see Sect.~\ref{sec:resultsEplusCMB} for details).}
    \label{fig:all_res_addCMBlens}
\end{figure*}

As can be seen in Fig.~\ref{fig:all_res_addCMBlens}, the combination and cross-correlation of \Euclid probes with CMB lensing noticeably improves the constraints on the nuisance parameters, in particular the ones related to intrinsic alignments in the pessimistic \Euclid scenario. There, the improvement on the parameters $\mathcal{A}_{\rm IA}$, $\eta_{\rm IA}$, and $\beta_{\rm IA}$ ranges, respectively, from 2.7\,\%, 1.9\,\%, and 5.1\,\% to 9.5\,\%, 7.7\,\%, and 15.6\,\% across all considered cosmological models. Those improvements stem from the kernel overlap between CMB lensing and galaxy WL; since the CMB lensing measurement is not affected by IA, its combination and cross-correlation with WL probes allows us to break degeneracies introduced by the IA nuisance parameters, incidentally improving the constraints on those parameters. A similar effect can be observed for the constraints on galaxy bias parameters, with improvements reaching up to 83\,\%. The improvements are especially seen on the high-redshift bias parameters in the \Euclid optimistic case; the cross-correlation of galaxy density and CMB lensing is probing the peak of the CMB lensing kernel ($z\sim2$) hence maximising its impact in the analysis at those redshifts. Unsurprisingly, the combination of \Euclid with \Planck observables is the least constraining. This comes from the fact that the CMB lensing information, dominant at small angular scales, cannot be optimally retrieved from \Planck's low-resolution data. On the other hand, the lensing data obtained by the more advanced CMB facilities will provide more information to counteract the degeneracies introduced by intrinsic alignment.

For cosmological parameters, the joint analysis with lensing from SO -- and even more so from a CMB-S4-like survey -- improves the constraints the most as expected, especially in comparison to the \Euclid pessimistic-only results. Parameters of the extended models particularly benefit from the combination, with the uncertainties on curvature being reduced by a factor of 3.7 at most, while the constraints on the MG parameter $\gamma$ can be improved by 50\,\%. DE parameters ($w_0$, $w_a$) are overall affected to a somewhat lesser extent, but still reach improvements of up to 80\,\%, cf.\ $w_a$ in the non-flat case compared to the pessimistic \Euclid-only case. The standard set of cosmological parameters shows an improvement mainly when a CMB-S4-like survey is considered, with an average of 11\,\% (maximum 44\,\%) for pessimistic \Euclid and 9\,\% (maximum 25\,\%) for the optimistic case, while other CMB scenarios show a below 5\,\% average improvement.

\subsection{Full \ExC joint analysis}\label{sec:resultsEcrossCMB}

\begin{figure*}
    \centering
    \includegraphics[width=\textwidth]{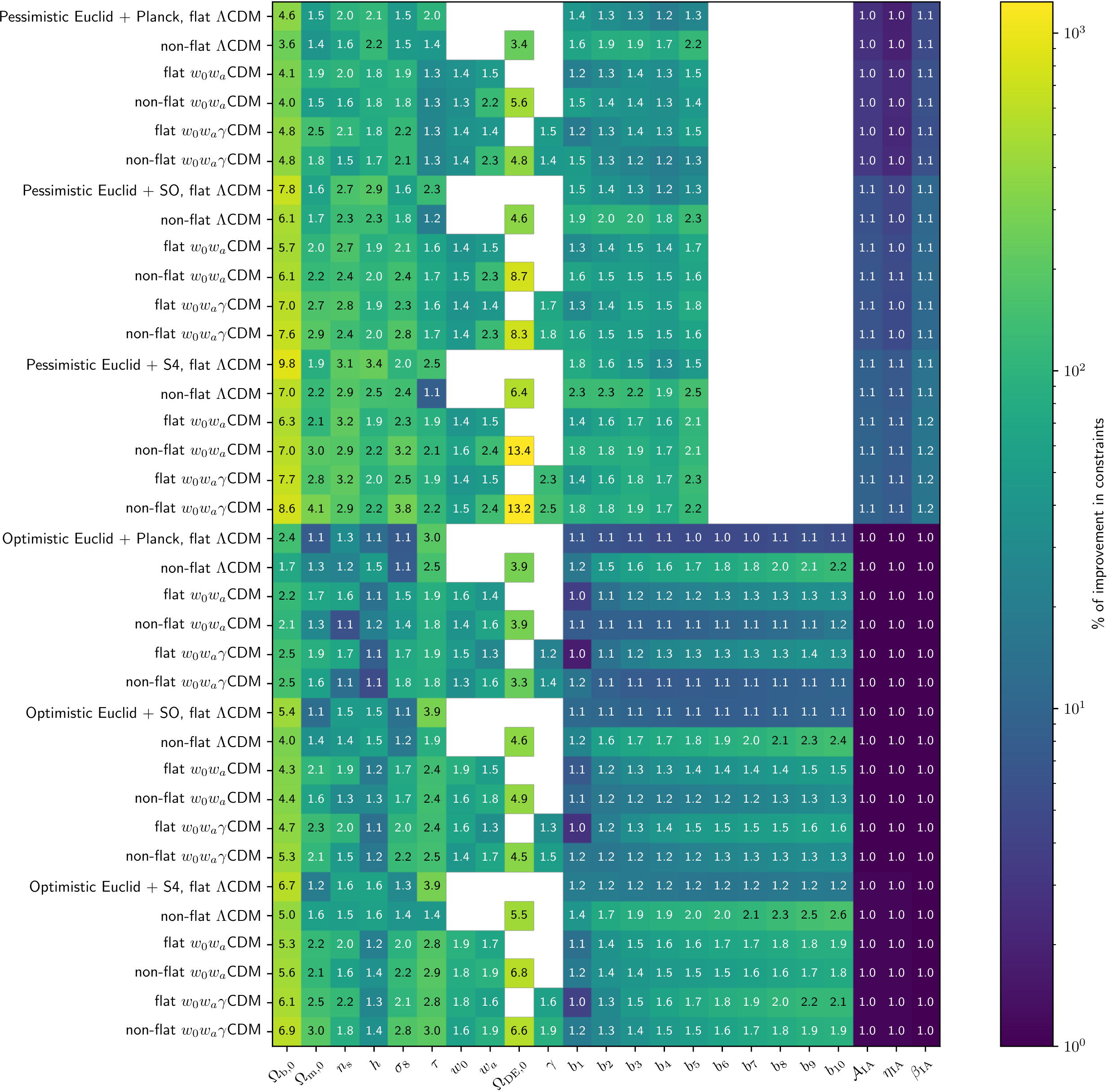}
    \caption{Predicted improvements on parameter constraints after adding all CMB probes to \Euclid, including all cross-correlations, for all scenarios (cosmological models and survey specifications) considered in the present study. Conventions are identical to those of Fig.~\ref{fig:all_res_addCMBlens}. We note that, for the optical depth $\tau$, since \Euclid alone is not able to constrain this parameter, we show instead the predicted improvement after adding \Euclid to CMB data.}
    \label{fig:all_res_addfullCMB}
\end{figure*}

In this second part of our analysis, we take into account all CMB probes (temperature, polarisation and lensing) and all their cross-correlations with \Euclid in the data vector of the Fisher analysis. Figure~\ref{fig:all_res_addfullCMB} summarises our forecasts for the full \ExC joint analysis by showing the improvements in constraints compared to the case where only \Euclid probes are considered.

The addition of CMB probes significantly improves the constraints on cosmological parameters in all the cosmological cases considered. We observe that in most scenarios, $\Omega_{\rm b, 0}$ and $\Omega_{\rm DE, 0}$ are the parameters that are best improved as a result of the joint analysis. In the case of $\Omega_{\rm b, 0}$, the factor of improvement across the different cosmological models ranges from 1.7 when \Planck is added to optimistic \Euclid results, and up to 9.8 when CMB-S4-like data are combined with a more pessimistic scenario for \Euclid (with an overall average factor of 5.3). This improvement is likely due to the fact that the shape of the CMB power spectrum and relative amplitudes of the acoustic peaks are highly sensitive to baryon density. For $\Omega_{\rm DE, 0}$, the improvement factor across cosmological models ranges between 3.3 and 13.4 (average 6.2, roughly twice as constraining as the addition of CMB lensing alone), echoing the constraining power of the CMB on curvature. On the contrary, the Hubble parameter $h$ is among the ones showing the least improvement, with an average factor of improvement of 1.7 (maximum 3.4). This indicates that the \Euclid main probes, GC and WL, are already powerful at constraining the background evolution of the Universe, and so CMB data do not add much more information. Moreover, CMB observables depend on $h$ mostly through the location of the acoustic peaks, namely the angular size of the sound horizon at recombination, which is an integrated quantity and only directly related to $h$ in the simplest models. Thus, the introduction of additional cosmological parameters -- also entering the computation of the angular size -- induces degeneracies that further reduce the constraining power of the CMB on $h$. Constraints on the other parameters of extensions to the baseline cosmological model, namely $w_0$, $w_a$ and $\gamma$, show a (relatively) moderate improvement with respect to the \Euclid-alone constraints, with an average factor of 1.6 (maximum 2.5). The full joint analysis with CMB provides on average an additional improvement factor of 1.4 (maximum 1.9) on these parameter constraints compared to the gains from adding CMB lensing alone.

Lastly, we underline the special status of one of the parameters considered in Fig.~\ref{fig:all_res_addfullCMB}, namely the optical depth of reionisation. Indeed, since the late-Universe probes of \Euclid are not sensitive to $\tau$ (and thus cannot constrain it) we show instead in the corresponding column the converse, namely the predicted improvement from the joint \ExC analysis compared to CMB-only constraints on $\tau$. The resulting gains range from a very modest 1.09 to an impressive 3.9 factor (average 2.1). The trends in those improvements are easily understood: the optimistic \Euclid scenario has a larger effect than the pessimistic one, and the worse the CMB specs are, the greater the amplitude of the improvement. These results originate mostly from the degeneracy between $\tau$ and the amplitude of perturbations $A_{\rm s}$ in CMB studies, which is broken when adding the tight constraints on $\sigma_8$ from large-scale surveys. The case of $\tau$ illustrates the two-way nature of the gains expected from the \ExC joint analysis, and means that we expect to break other CMB-related degeneracies in the future (e.g. the ones involving the tensor-to-scalar ratio), as well as probe-specific systematic effects.

Overall, we find that the standard cosmological model and its extensions benefit greatly from the joint analysis of \Euclid and the CMB, in terms of precision on the measured parameters. It also appears quite clear that, as was to be expected, the addition of CMB constraints is more helpful for the pessimistic \Euclid scenario than for the optimistic one. Finally, we note that a CMB-S4-like experiment is obviously the source of the highest improvements when combined with \Euclid results, given its superior resolution and sensitivity.

\begin{figure}
    \centering
    \includegraphics[width=\columnwidth]{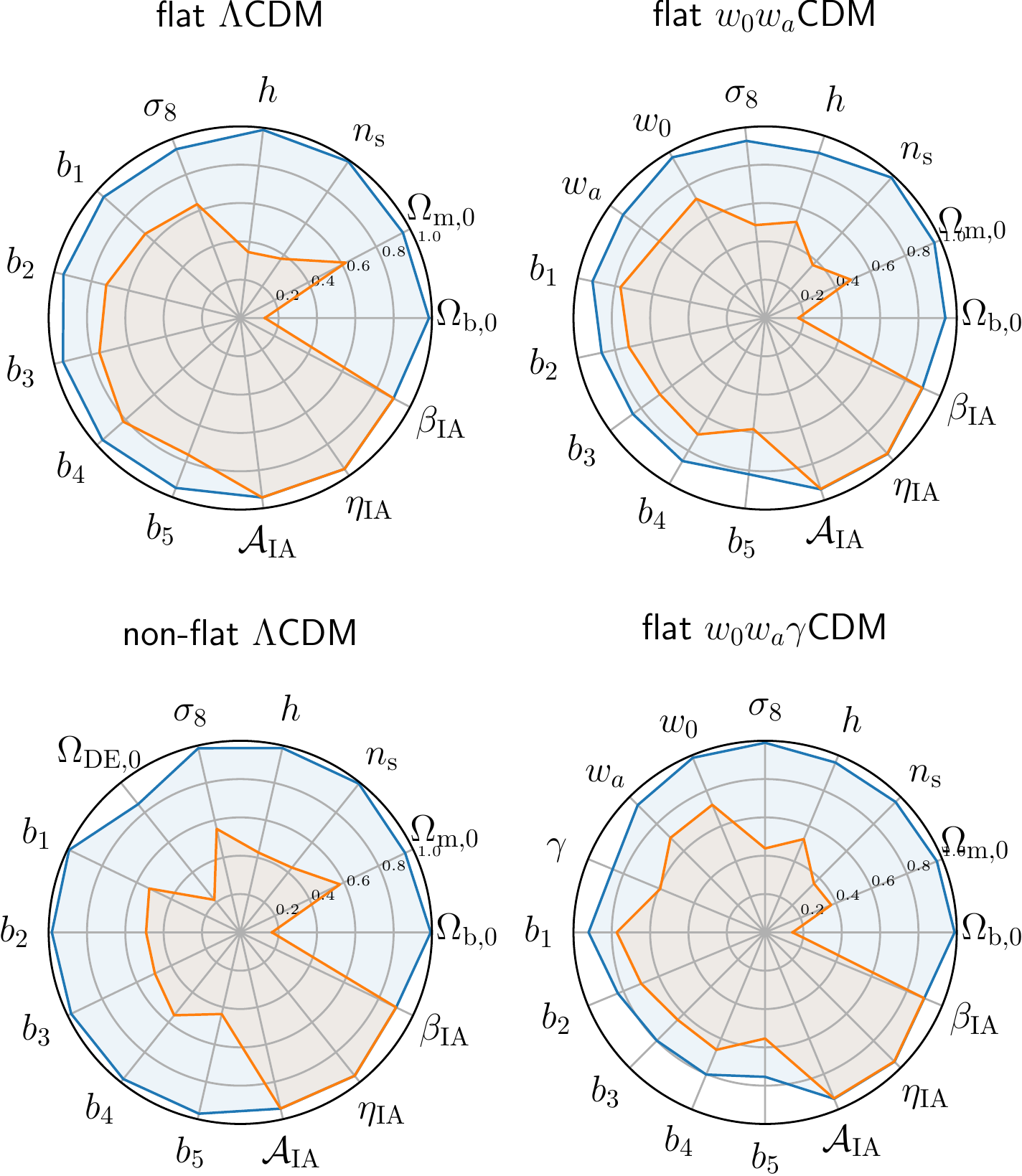}
    \caption{Ratio of predicted $1\,\sigma$ uncertainties (see end of Sect.~\ref{sec:resultsEcrossCMB}) showing how constraints are tightened after adding CMB lensing (blue) or all CMB probes (orange) when compared to the \Euclid-only constraints (black outer rim), assuming a pessimistic \Euclid scenario and SO-like CMB data, for four selected cosmological models (from top to bottom, left to right: flat $\Lambda$CDM; flat $w_0$$w_a$CDM; non-flat $\Lambda$CDM; and flat $w_0$$w_a$$\gamma$CDM)}.
    \label{fig:radar_plot}
\end{figure}

\begin{figure}
    \centering
    \includegraphics[width=\columnwidth]{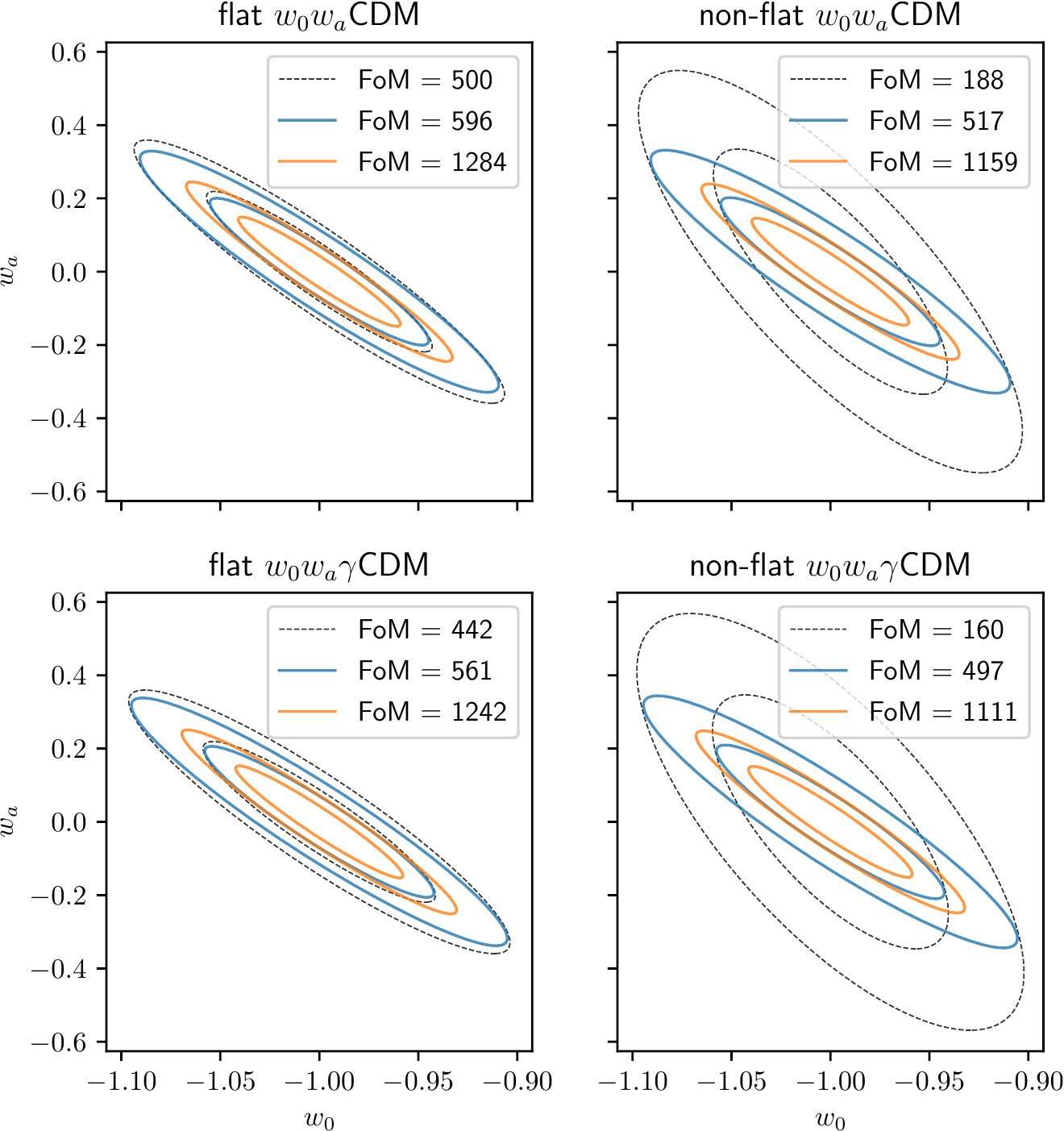}\\
    \includegraphics[width=\columnwidth]{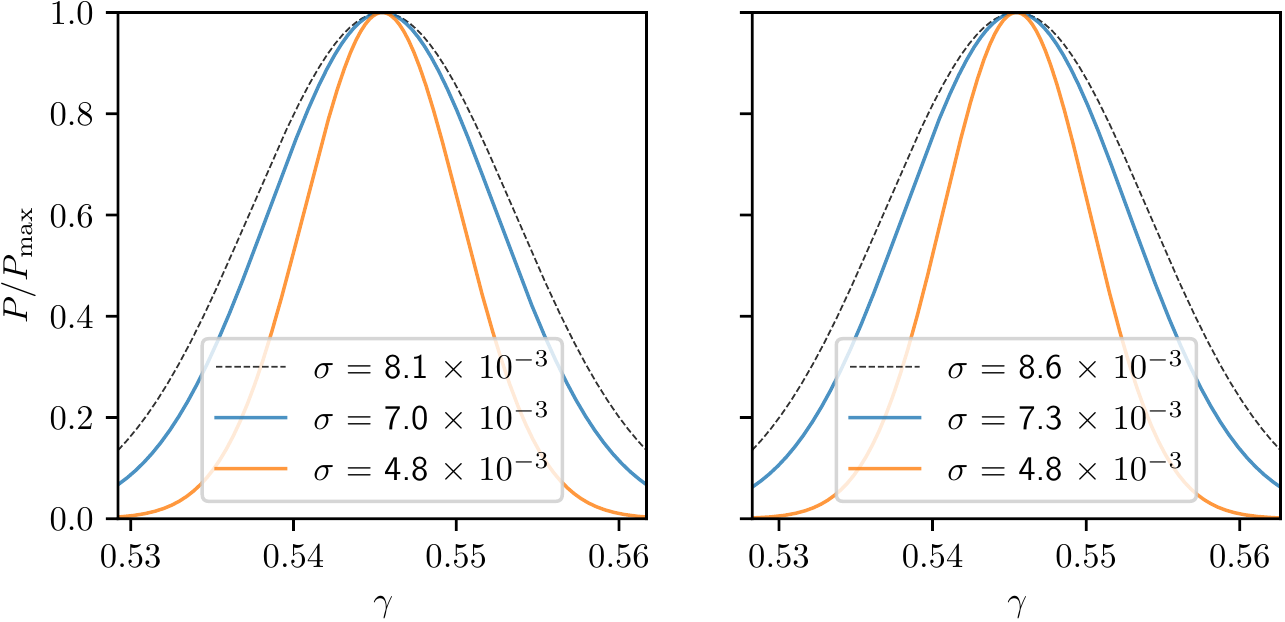}
    \caption{Predicted 1D and 2D marginalised distributions for the $w_0$--$w_a$ and/or $\gamma$ parameters, for four cosmological model: flat $w_0$$w_a$CDM (top left); non-flat $w_0$$w_a$CDM (top right); flat $w_0$$w_a$$\gamma$CDM (middle and bottom left); and non-flat $w_0$$w_a$$\gamma$CDM (middle and bottom right). This is for the same scenario as Fig.~\ref{fig:radar_plot}, using the same colour coding (where \Euclid-only constraints are here in dashed black). Each plot is accompanied by the corresponding FoM or $1\,\sigma$ uncertainties.}
    \label{fig:w0wagamma_plot}
\end{figure}

For further illustration purposes, Fig.~\ref{fig:radar_plot} highlights a particular scenario from Figs.~\ref{fig:all_res_addCMBlens} and \ref{fig:all_res_addfullCMB}, namely an SO-like CMB survey combined with a pessimistic \Euclid survey, for four of our six cosmological models. The results are presented in the form of `radar' plots, which show this time the ratio of uncertainties
\begin{equation}
    \frac{\sigma_{\rm after}}{\sigma_{\rm before}} \,,
\end{equation}
namely the inverse of the ratios shown in Figs.~\ref{fig:all_res_addCMBlens} and \ref{fig:all_res_addfullCMB}. The distance from the centres of these plots is a visual representation of the $1\,\sigma$ uncertainty on all parameters of our analysis, where a length of one corresponds to the \Euclid-only constraints. We observe here once again how adding CMB lensing information (blue lines) affects mostly the nuisance parameters and extended models parameters, whereas the addition of all CMB probes has a more dramatic overall effect on all parameters for all models. Some additional visualisation of our results is presented in Fig.~\ref{fig:w0wagamma_plot}: for the same aforementioned scenario (pessimistic \Euclid $+$ SO), we plot the 2D marginalised constraints in the $w_0$--$w_a$ plane, and the 1D marginalised distribution for the MG parameter $\gamma$. Each plot contains in its legend either the corresponding `figures of merit' (FoM, proportional to the inverse of the area of the ellipse of constraints, see \citetalias{Euclid_IST} for an exact definition) in the case of $w_0$--$w_a$ constraints, or the $1\,\sigma$ uncertainties for $\gamma$. The FoM figures are particularly relevant because they represent one of the main expected results of \Euclid, namely constraints on the equation of state of the elusive DE. Our plots illustrate in particular how, even though CMB data by themselves do not provide much constraints on the DE equation of state, the improvements on the determination of other parameters can still yield a significant shrinkage of the 2D contours, thanks to correlations between parameters. The effect is particularly visible for the non-flat models, where the substantial CMB constraints on curvature play a large role in the improvements, even when only adding CMB lensing data. The full numerical results for this particular scenario are shown in Table~\ref{tab:resEuclidxCMB}.

\begin{table*}[h!]
    \caption{Predicted constraints on cosmological parameters from the joint \ExC analysis.}
    \centering
    {\renewcommand{\arraystretch}{1.2}
    \begin{tabular}{rccccccccc}
        \hline
        Model & $\Omega_{\rm b, 0}$ & $\Omega_{\rm m, 0}$ & $n_{\rm s}$ & $h$ & $\sigma_{8}$ & $\Omega_{\rm DE,0}$ & $w_0$ & $w_a$ & $\gamma$ \\
        \hline
        \hline
        & \multicolumn{9}{c}{\underline{After combination with CMB lensing}} \\
        flat $\Lambda$CDM & 0.025 & 0.0062 & 0.0052 & 0.0036 & 0.0029 & \dots & \dots & \dots & \dots \\
        non-flat $\Lambda$CDM & 0.026 & 0.0062 & 0.0054 & 0.0042 & 0.0032 & 0.0085 & \dots & \dots & \dots \\
        flat $w_0$$w_a$CDM & 0.029 & 0.010 & 0.0055 & 0.0041 & 0.0042 & \dots & 0.036 & 0.13 & \dots \\
        non-flat $w_0$$w_a$CDM & 0.030 & 0.011 & 0.0056 & 0.0044 & 0.0046 & 0.0096 & 0.037 & 0.13 & \dots \\
        flat $w_0$$w_a$$\gamma$CDM & 0.038 & 0.015 & 0.0057 & 0.0045 & 0.0049 & \dots & 0.038 & 0.14 & 0.013 \\
        non-flat $w_0$$w_a$$\gamma$CDM & 0.037 & 0.014 & 0.0058 & 0.0046 & 0.0050 & 0.012 & 0.038 & 0.14 & 0.013 \\
        \hline
        & \multicolumn{9}{c}{\underline{Full \ExC joint analysis}} \\
        flat $\Lambda$CDM & 0.0032 & 0.0040 & 0.0020 & 0.0012 & 0.0019 & \dots & \dots & \dots & \dots \\
        non-flat $\Lambda$CDM & 0.0043 & 0.0038 & 0.0024 & 0.0018 & 0.0018 & 0.0022 & \dots & \dots & \dots \\
        flat $w_0$$w_a$CDM & 0.0054 & 0.0053 & 0.0021 & 0.0024 & 0.0022 & \dots & 0.027 & 0.099 & \dots \\
        non-flat $w_0$$w_a$CDM & 0.0050 & 0.0049 & 0.0024 & 0.0023 & 0.0019 & 0.0029 & 0.026 & 0.096 & \dots \\
        flat $w_0$$w_a$$\gamma$CDM & 0.0055 & 0.0056 & 0.0021 & 0.0025 & 0.0022 & \dots & 0.028 & 0.10 & 0.0088 \\
        non-flat $w_0$$w_a$$\gamma$CDM & 0.0051 & 0.0052 & 0.0024 & 0.0023 & 0.0020 & 0.0031 & 0.027 & 0.10 & 0.0088 \\
        \hline
    \end{tabular}
    }
    \tablefoot{We report here the predicted constraints on cosmological parameters from the joint analysis of a pessimistic \Euclid survey and SO-like CMB probes (CMB lensing only and full \ExC combination), expressed as the ratio of marginalised 1$\sigma$ uncertainties over their corresponding fiducial values (same conventions as Table~\ref{tab:resEuclid})}
    \label{tab:resEuclidxCMB}
\end{table*}


\section{Conclusions}
\label{sec:conclusions}

\Euclid will revolutionise our vision of the Universe by mapping its matter distribution and providing us with exquisite measurements of its main probes, namely WL and GC, hence probing the nature of the accelerated expansion. At the same time, a wealth of data will be available from large surveys ranging from optical (e.g. Vera Rubin Observatory) to radio (e.g. the Square Kilometre Array) and the millimetre (SO).

In this context, the combination and cross-correlation of the upcoming \Euclid and CMB survey data will be of prime importance for cosmological analyses. In preparation for this synergetic data exploitation, we have conducted a forecast analysis on the precision achievable on the most important cosmological parameters, including the ones describing DE and MG. To do so, we have used a standard Fisher formalism, as well as a posterior-fitting approach based on actual CMB data, assuming nominal specifications of the current and future surveys. For \Euclid, we have set up our forecasting pipeline in a consistent and complementary manner with respect to recently published forecasts on \Euclid-specific probes (\citetalias{Euclid_IST}) namely GC, WL, and their cross-correlation. In terms of CMB specifications, we have considered the cases of a \Planck-like survey, the SO, and the CMB-S4 experiment.

In our analysis, we derived constraints for a baseline model, the $\Lambda$CDM cosmological model with (minimal) massive neutrinos, described by six cosmological parameters. We also explored: a possible non-zero curvature; a redshift-dependent DE equation of state; and a phenomenological description of MG. In addition, we considered 13 nuisance parameters, consistent with the forecasts previously presented by the Euclid Consortium (\citetalias{Euclid_IST}). We have evaluated the expected impact of the combination and cross-correlation of \Euclid with CMB data in terms of improvement factors with respect to the pure \Euclid constraining capabilities.

We found that the addition of CMB data (and their cross-correlations with \Euclid observables) helps substantially in reducing the impact of nuisance parameters on cosmological constraints, leading for example to an improvement of up to 15\,\% on intrinsic alignment parameters and 2.6 times smaller uncertainties on galaxy bias. The joint \Euclid and CMB analysis also leads to a substantial improvement on all cosmological parameters of the standard $\Lambda$CDM model, but with varying amplitude, noting for example the higher gains for $\Omega_{\rm b,0}$, but the lower gains for the reduced Hubble parameter $h$. Depending on the cosmological model considered (e.g. when a time-varying DE equation of state, a non-zero curvature, or a phenomenological MG are added), the improvement in the constraints ranges from tens of percent to factors of a few. In particular, the overall improvement in the $w_0$--$w_a$ constraints for all three CMB experiments considered here implies an increase in the corresponding figure-of-merit by a factor two and up to almost seven in certain cases (cf. Fig.~\ref{fig:w0wagamma_plot}).

Finally, we stress that these results were obtained in configurations where the full complexity and the completeness of CMB and \Euclid observables was not modelled. Indeed, among the numerous `ingredients' that were not included, one can mention:
\begin{itemize}[topsep=0pt]
    \item uncertainties on the galaxy redshift distribution, as well as on the parameters of the photometric error modelling (all fixed to their fiducial values here);
    \item scale dependence of the galaxy bias, especially on non-linear scales;
    \item correlations between GCs and all other probes considered;
    \item BAO reconstruction as an additional probe extracted from the spectroscopic data;
    \item magnification bias and redshift-space distortions in the GCp analysis;
    \item additional non-Gaussian terms in the covariances (thus reducing the constraining power of the considered probes), for example due to non-linear effects or the super-sample covariance.
\end{itemize}
The inclusion of some of those points in the analysis could significantly improve the \Euclid-only constraints, while some could also severely degrade them. For these latter cases the combination with CMB data could significantly (and positively) affect the final results. Despite this, our results shown here in a more ideal setting not only confirm and highlight the benefits of combining and cross-correlating \Euclid with the CMB, but also show how powerful cross-correlations are for actual data analysis. On the basis of these results, future work will consider a more exhaustive set of non-standard cosmological models, and the validation of these forecasts by means of an end-to-end pipeline involving more realistic instrumental effects.


\begin{acknowledgements}
    \AckEC
    Marco Baldi acknowledges support by the project `Combining Cosmic Microwave Background and Large Scale Structure data: an Integrated Approach for Addressing Fundamental Questions in Cosmology', funded by the PRIN-MIUR 2017 grant 2017YJYZAH. Jose R. Bermejo-Climent is supported by an INAF fellowship under the agreement INAF-IAC. Mario Ballardini, Jose R. Bermejo-Climent, Fabio Finelli acknowledge financial contribution from the contract ASI/INAF for the Euclid mission n.2018-23-HH.0. DP, Fabio Finelli acknowledge financial support by ASI Grant 2016-24-H.0 and and the agreement n.2020-9-HH.0 ASI-UniRM2 `Partecipazione italiana alla fase A della missione LiteBIRD'. The SISSA group acknowledges support by the ASI/INAF contracts Euclid-IC (I/031/10/0), ASI-COSMOS (\url{cosmosnet.it}), ASI-LiteBIRD contracts, the INDARK INFN Initiative (\url{web.infn.it/CSN4/IS/Linea5/InDark}), and the MIUR PRIN 2015 grant `Cosmology and Fundamental Physics: illuminating the Dark Universe with Euclid'. Giulio Fabbian acknowledges the support of the European Research Council under the Marie Sk\l{}odowska Curie actions through the Individual Global Fellowship No.~892401 PiCOGAMBAS. Stéphane Ili\'c acknowledges financial support from the European Research Council under the European Union's Seventh Framework Programme (FP7/2007-2013)/ERC Grant Agreement No. 617656 `Theories and Models of the Dark Sector: Dark Matter, Dark Energy and Gravity. Louis Legrand acknowledges support from CNES's funding of the Euclid project and a SNSF Eccellenza Professorial Fellowship (No. 186879). Domenico Marinucci acknowledges support from the MIUR Excellence Project awarded to the Department of Mathematics, Università di Roma Tor Vergata, CUP E83C18000100006. Marina Migliaccio was supported by the program for young researchers `Rita Levi Montalcini' year 2015. Alessandro Renzi was supported by the project `Combining Cosmic Microwave Background and Large Scale Structure data: an Integrated Approach for Addressing Fundamental Questions in Cosmology', funded by the MIUR Progetti di Rilevante Interesse Nazionale (PRIN) Bando 2017 - grant 2017YJYZAH; and acknowledges funding from Italian Ministry of Education, University and Research (MIUR) through the `Dipartimenti di eccellenza' project Science of the Universe. Stefano Camera acknowledges support from the `Departments of Excellence 2018-2022' Grant (L.\ 232/2016) awarded by the Italian Ministry of University and Research (\textsc{mur}). Matteo Martinelli acknowledges support from the Centro de Excelencia Severo Ochoa Program SEV-2016-059 and from `la Caixa' Foundation (ID 100010434), with fellowship code LCF/BQ/PI19/11690015. Alkistis Pourtsidou is a UK Research and Innovation Future Leaders Fellow, grant MR/S016066/1. Ziad Sakr acknowledge support from the IRAP and IN2P3 Lyon computing centers. Domenico Sapone acknowledges financial support from Fondecyt Regular project number 1200171. Isaac Tutusaus acknowledges support from the Spanish Ministry of Science, Innovation and Universities through grant ESP2017-89838, and the H2020 programme of the European Commission through grant 776247. Victoria Yankelevich acknowledges funding from the European Research Council (ERC) under the European Union's Horizon 2020 research and innovation programme (grant agreement No. 769130).
\end{acknowledgements}

\bibliographystyle{aa}
\bibliography{References}


\begin{appendix}

\section{Fitted-Fisher approach}
\label{sec:Appendix}

\subsection{Method}

We describe in this subsection a method -- thereafter referred to as the `fitted-Fisher approach' -- designed to combine a theoretical \Euclid Fisher matrix with constraints from the actual CMB data and likelihood, more specifically in the context of \Planck.

For models that are well constrained by \Planck data -- such as the minimal $\Lambda$CDM -- the posterior distribution of cosmological parameters follows a multivariate Gaussian quite closely. One can thus summarise the constraints from the \Planck data by extracting the mean and the covariance matrix of cosmological parameters directly from the samples of a previously computed Monte Carlo Markov chain. The inverse of such a matrix can then be interpreted as a `fitted-Fisher matrix', which can then be added to other `actual' Fisher matrices in order to emulate the addition of CMB constraints.

For models that are less constrained (e.g. $w$CDM, $w_0w_a$CDM), one can fit the posterior distribution of the MCMC approximately with a multi-dimensional Gaussian and thus obtain the covariance matrix. This fit is performed in the vicinity of the \Euclid fiducial values of the extra parameters (e.g. $w_0=-1$, $w_a=0$, etc). This ensures that the resulting mean, covariance and fitted-Fisher matrix most accurately reflect what the actual \Euclid $+$ \Planck combination would yield if we assume that the true underlying model is the fiducial one chosen for the \Euclid Fisher matrix.

A technical point in this approach is that one cannot in practice directly combine the \Euclid Fisher matrix and the \Planck fitted-Fisher matrix described above, since they are not `evaluated' at the same point in the parameter space -- the \Euclid Fisher matrix is evaluated around an arbitrarily chosen fiducial point, whereas by construction the CMB fitted-Fisher matrix is evaluated around the point corresponding to the maximum of the \Planck likelihood. As a consequence, some care has to be taken when combining the two matrices, and the resulting combination has a third, different effective fiducial point. In practice, since the separation between the two fiducial points is small, we decided here to simply shift the \Planck fitted-Fisher matrix to the \Euclid fiducial point, thus simplifying the addition of the two sets of constraints.

We note that, by construction, this fitted-Fisher approach is incapable of incorporating the impact of any cross-correlation between CMB and \Euclid observables, since they are being considered as purely independent from each other. This approach and its results are thus presented here as a study case in the broader context of the full \ExC combination, which traditional Fisher forecasts can assess.
\subsection{Results and discussion}

\begin{figure*}
    \centering
    \includegraphics[width=\textwidth]{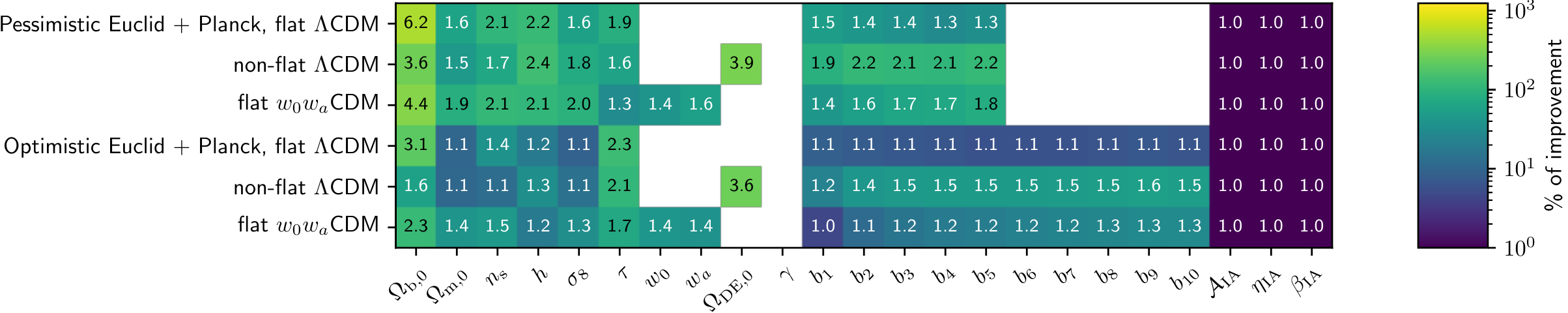}
    \caption{Predicted improvements on parameter constraints after adding \Planck CMB data to \Euclid via the fitted-Fisher approach. Conventions are identical to Fig.~\ref{fig:all_res_addCMBlens} and with the same remark for $\tau$ as in Fig.~\ref{fig:all_res_addfullCMB}.}
    \label{fig:fitted-fish}
\end{figure*}

The results from the \Euclid and real \Planck data combination via the fitted-Fisher approach are summarised in Fig.~\ref{fig:fitted-fish}. We see a remarkable agreement when compared to the traditional Fisher formalism (cf.\ Fig.~\ref{fig:all_res_addfullCMB}); the two methods show similar trends and orders of magnitude for factors of improvements, across all data combinations and cosmological models considered. We note that, since we fitted the available posteriors of publicly available \Planck chains,\footnote{\url{pla.esac.esa.int/pla}} the analysis is restricted to only three of our six models (namely flat $\Lambda$CDM, non-flat $\Lambda$CDM and flat ${w_0}{w_a}$CDM). We acknowledge several causes that could explain the observed (relatively) small differences between the two methods, including the fact that the fitted-Fisher approach does not account for the covariance between \Planck and \Euclid (thus potentially overestimating some constraints) and the fact that our effective approach in defining the specifications of the \Planck-like survey in the traditional Fisher approach may result in non-trivial deviations from the true \Planck survey characteristics.

\end{appendix}

\end{document}